\documentclass[sigconf]{acmart}
\pdfoutput=1
\usepackage[utf8]{inputenc}
\usepackage{graphicx} 
\usepackage{tabularx}
\usepackage{float}
\usepackage{colortbl}
\usepackage{array}
\usepackage{caption}
\usepackage{fancyhdr}
\usepackage{tcolorbox}
\usepackage{subcaption} 
\usepackage{booktabs}
\usepackage[table]{xcolor} 
\usepackage{longtable}

\tcbuselibrary{breakable}

\AtBeginDocument{%
  }

\copyrightyear{2025}
\acmYear{2025}
\setcopyright{cc}
\setcctype{by-nc-sa}
\acmConference[CUI '25]{Proceedings of the 7th ACM Conference on Conversational User Interfaces}{July 8--10, 2025}{Waterloo, ON, Canada}
\acmBooktitle{Proceedings of the 7th ACM Conference on Conversational User Interfaces (CUI '25), July 8--10, 2025, Waterloo, ON, Canada}
\acmDOI{10.1145/3719160.3736623}
\acmISBN{979-8-4007-1527-3/2025/07}

\begin{document}

\title{NeuroChat: A Neuroadaptive AI Chatbot for Customizing Learning Experiences}


\author{Dünya Baradari}
\affiliation{%
  \institution{MIT Media Lab}
  \city{Cambridge, MA}
  \country{United States}}

\author{Nataliya Kosmyna}
\affiliation{%
  \institution{MIT Media Lab}
  \city{Cambridge, MA}
  \country{United States}}

  \author{Oscar Petrov}
\affiliation{%
  \institution{Brown University}
  \city{Providence, RI}
  \country{United States}}

    \author{Rebecah Kaplun}
\affiliation{%
  \institution{MIT Media Lab}
  \city{Cambridge, MA}
  \country{United States}}

  \author{Pattie Maes}
\affiliation{%
  \institution{MIT Media Lab}
  \city{Cambridge, MA}
  \country{United States}}

\renewcommand{\shortauthors}{Baradari, Kosmyna, Petrov, Kaplun, and Maes}

\begin{abstract}
Generative AI is reshaping education by enabling personalized, on-demand learning experiences. However, current AI systems lack awareness of the learner’s cognitive state, limiting their adaptability. In parallel, electroencephalography (EEG)-based neuroadaptive systems have shown promise in enhancing engagement through real-time physiological feedback. This paper introduces \textit{NeuroChat}, a neuroadaptive AI tutor that integrates real-time EEG-based engagement tracking with a large language model to adapt its conversational responses. By continuously monitoring learners’ cognitive engagement, NeuroChat dynamically adjusts content complexity, tone, and response style in a closed-loop interaction. In a within-subjects study ($n=24$), NeuroChat significantly increased both EEG-measured and self-reported engagement compared to a non-adaptive chatbot. However, no significant differences in short-term learning outcomes were observed. These findings demonstrate the feasibility of real-time brain–AI interaction for education and highlight opportunities for deeper personalization, longer-term adaptation, and richer learning assessment in future neuroadaptive systems.
  
\end{abstract}



\begin{teaserfigure}
    \centering
    \begin{minipage}{0.95\textwidth}
        \centering
        \includegraphics[width=\linewidth]{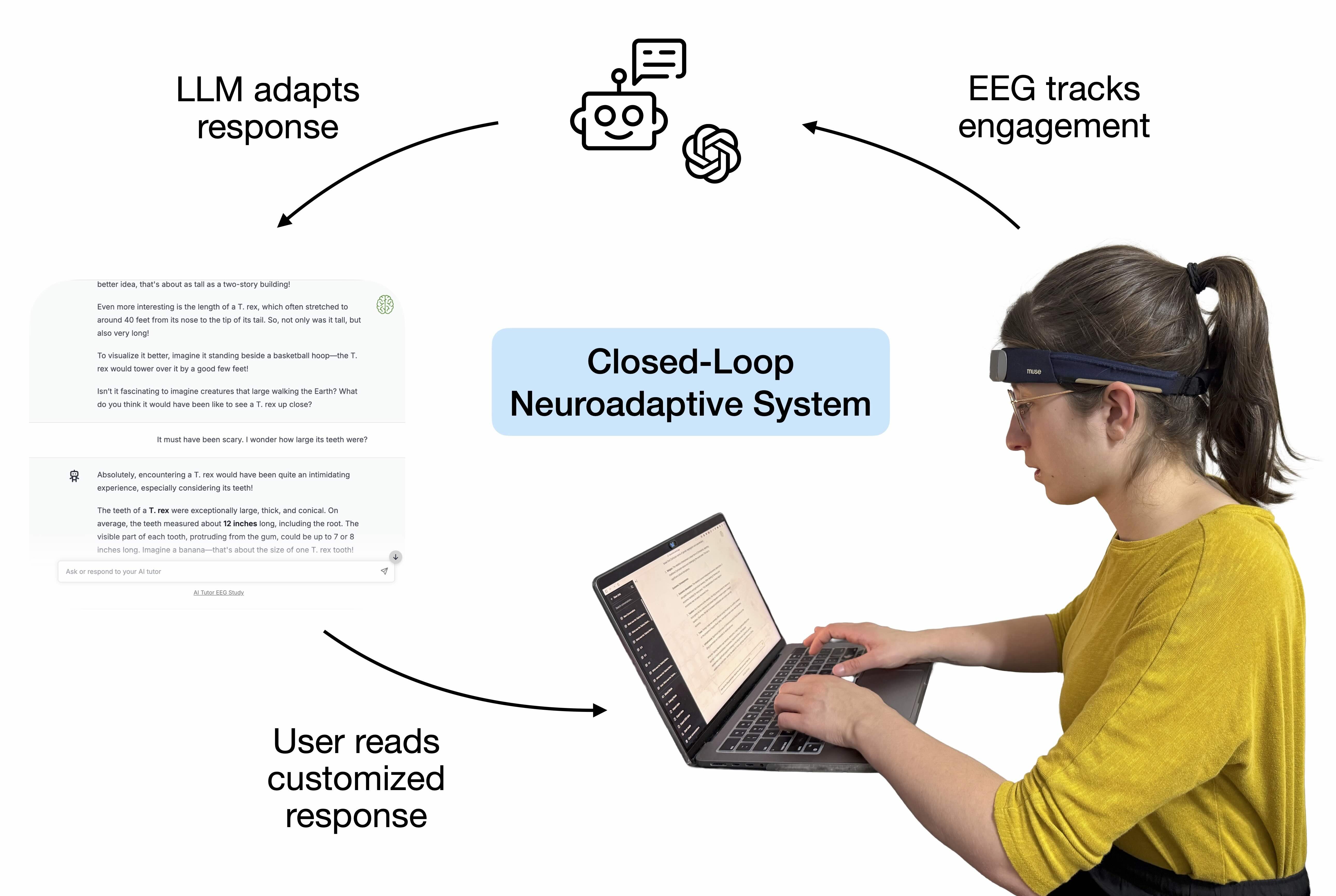}
    \end{minipage}
\caption{Overview of the NeuroChat neuroadaptive LLM system. A wearable dry-electrode EEG headband collects data from the brain and sends it to the NeuroChat web app, which computes the user’s level of engagement. The engagement score is sent with each request to the LLM, allowing it to adapt its response style to the user in real time.}
\Description{A diagram showing the NeuroChat closed-loop neuroadaptive system. A user wears a Muse EEG headband and looks at a laptop screen showing chatbot responses. The system forms a loop where EEG tracks engagement, sends data to a large language model (LLM), which adapts its response, and the user reads the response.}
    \label{fig:teaser}
\end{teaserfigure}


\maketitle

\section{Introduction}

Generative Artificial Intelligence (AI) is reforming the educational landscape, as Large Language Models (LLMs) present new opportunities for personalized, on-demand learning. AI-powered tutors, such as ChatGPT Edu \cite{OpenAI2024-bj} and Khanmigo \cite{KhanAcademy2025-on}, are being integrated into educational settings, pointing towards a future in which learners can interact dynamically with AI, receive customized explanations, and engage in self-directed inquiry. AI-generated content has been shown to improve learning motivation \cite{Leong2024-hu} and enhance teaching efficiency \cite{Mulyani2025-sj}, as educators leverage these tools to tailor lesson plans, streamline administrative tasks, and develop adaptive instructional materials \cite{The-Open-Innovation-Team-and-Department-for-Education2024-cs}. At the same time, the opening of AI-only schools in the United Kingdom and the United States \cite{David-Game-College2024-gh, Schultz2025-on} demonstrates the growing acceptance of AI-driven adaptive learning platforms. These systems claim to dynamically adjust educational content to individual learners’ strengths and weaknesses, promising a level of personalization that traditional classroom settings often struggle to achieve \cite{Graf2009-ts}. 

Despite this enthusiasm surrounding the integration of LLMs into education, several unresolved challenges persist. Key issues include AI-generated hallucinations and concerns regarding data privacy \cite{Kasneci2023-id}. Additionally, cognitive offloading, where students become overly reliant on AI for information, remains a significant concern, as it may hinder the development of independent research and critical reasoning skills. Indeed, research shows that while LLM-based chatbots can enhance engagement \cite{Heung2025-xu}, foster critical thinking skills \cite{Essel2024-oa}, and improve academic performance \cite{Deng2025-uk}, excessive reliance on them for cognitive tasks can lead to the erosion of essential skills \cite{Dubey2024-md}. 

Therefore, it is crucial that adaptive learning systems appropriately challenge learners without overwhelming them, ideally guiding them toward the Zone of Proximal Development. Unlike human teachers, however, LLMs lack the ability to assess a learner’s cognitive and attentional state without explicit communication by the learner. As a result, LLMs may inadvertently overload students with excessive cognitive demands, present materials at inappropriate difficulty levels, or fail to recognize fluctuations in engagement. These shortcomings could ultimately hinder the effectiveness of learning outcomes.

Meanwhile, research in neuroadaptive learning systems has demonstrated that real-time physiological feedback, particularly from electroencephalography (EEG), can significantly enhance engagement and cognitive adaptation. EEG-based brain-computer interfaces (BCIs) can measure engagement, attention, and cognitive load, dynamically adjusting learning materials according to a learner’s real-time neural state \cite{Kosmyna2019-xi, Parsons2021-js}. Several closed-loop EEG systems have been developed to optimize learning by modulating content complexity based on neurophysiological signals. For example, the BRAVO system \cite{Marchesi2013-rg} detects fluctuations in attention and adjusts e-learning materials accordingly, while Thinking Cap \cite{Mills2017-au} integrates EEG-based cognitive load assessment into an Intelligent Tutoring System (ITS), modifying text complexity based on engagement levels. Similarly, FOCUS \cite{Huang2014-ch} adapts learning materials for children by integrating EEG-driven interventions during reading sessions. These studies show that adaptive educational environments, when informed by physiological feedback, can improve learning retention and engagement.

However, existing neuroadaptive systems are currently constrained by pre-scripted content, limiting adaptations to predefined difficulty levels rather than dynamically generating new instructional materials. For example, the Brain Automated Chorales (BACh) system \cite{Yuksel2016-ae} offers only four fixed difficulty levels for a piano piece, selecting among these based on the user’s cognitive load, which restricts the flexibility and responsiveness of the learning experience. This limitation raises an important question: Can we integrate real-time neuroadaptive EEG feedback with the content-generation capabilities of LLMs to create an AI tutor that adapts to a learner’s cognitive state?

A truly adaptive learning system should not only be capable of generating adaptive content but also of recognizing engagement levels and cognitive load—an ability that is fundamental to effective teaching \cite{Klieme2006-uq}. Pedagogical theories such as Teaching at the Right Level (TaRL) \cite{The-Abdul-Latif-Jameel-Poverty-Action-Lab-J-PAL-2022-nq} and Cognitive Load Theory (CLT) \cite{de-Jong2010-ff} emphasize that cognitive adaptation is crucial for learning success. When educational tools present material that exceed a learner’s cognitive capacity, learning is significantly hindered \cite{Fisch2017-zd}. Research has shown that maintaining optimal cognitive load enhances knowledge retention and critical thinking skills \cite{Yuksel2016-ae}, yet current AI tutors lack mechanisms to assess and adjust for cognitive strain in real-time. 

To address this challenge, we introduce NeuroChat, a real-time\footnote{Here, “real-time” refers to continuous, low-latency updates suitable for interactive systems, with engagement computed at 1 Hz using a sliding window to ensure stability without sacrificing responsiveness.} neuroadaptive AI tutor that integrates EEG-based engagement tracking with LLM-driven content generation. NeuroChat continuously monitors EEG-derived engagement levels and dynamically adjusts the depth, complexity, and style of content based on a learner’s cognitive state (Figure \ref{fig:teaser}). This work makes the following contributions:

\begin{enumerate}
    \item A novel integration of neuroadaptive learning with generative AI, creating a closed-loop adaptation mechanism where real-time EEG data informs and modifies interactions with an LLM tutor to optimize cognitive load and engagement.
    \item High accessibility and usability, achieved by implementing a lightweight, browser-based, wearable system with dry-electrode EEG headbands, ensuring usability beyond laboratory settings.
    \item Empirical evaluation of NeuroChat’s effectiveness, examining how real-time neuroadaptive AI tutoring impacts learning outcomes, cognitive engagement, and user experience compared to non-adaptive AI tutoring.
\end{enumerate}

\section{Related Work}

\subsection{Engagement in Learning}

\subsubsection{Defining and Conceptualizing Engagement}

Engagement is a widely used term in education and psychology, though its definition varies across disciplines \cite{Azevedo2015-ku}. In educational settings, the term can be conceptualized as a ‘multidimensional construct encompassing behavioral, emotional, and cognitive dimensions,’ \cite{Fredricks2004-mn} which has been shown to be linked to positive learning outcomes, including increasing student motivation \cite{Fredricks2004-mn, Steinmayr2019-vu}. Expanding on this, Reeve and Tseng \cite{Reeve2011-vd} introduce a fourth dimension, agentic engagement, to describe students’ contributions to their learning experience. From the perspective of flow theory, learner engagement can be enhanced by designing learning activities that promote autonomy and provide appropriate challenges to learners’ skill level \cite{Shernoff2014-ys}. Sinatra et al. \cite{Sinatra2015-nt} further distinguish between microlevel engagement, which refers to moment-to-moment cognitive focus on a task, and macro-level engagement, which applies to larger social and educational contexts, such as classrooms or institutions. Micro-level engagement can be assessed through physiological techniques such as brain imaging, skin conductivity, or eye tracking, whereas macrolevel engagement is typically measured through sociocultural analysis, observations, or ratings.

In cognitive neuroscience, engagement is closely linked to sustained attention and tonic alertness, reflecting a person’s sustained cognitive effort \cite{Oken2006-gr}. However, engagement extends beyond attention, incorporating factors such as intrinsic motivation and task involvement \cite{Kamzanova2011-ff, Reinerman2006-df}. Unlike cognitive load, which reflects the mental demands on working memory, engagement captures both effort and motivation in a task-driven context. For this study, we define engagement as the sustained allocation of cognitive and attentional resources toward a task, influenced by motivation and mental effort.

\subsubsection{Physiological Measures of Engagement}

Various physiological technologies have been explored to assess engagement in digital learning environments, including video analysis, eye tracking, and biosensors. Classroom video analysis has been used to monitor student attention, with Raca and Dillenbourg \cite{Raca2013-dl} utilizing video recordings and later incorporating a computer vision model to approximate eye gaze \cite{Raca2015-es}. However, these models have limited accuracy in estimating attention levels, as they attempt to infer complex cognitive states from external behavioral approximations that are often ambiguous and context-dependent. Eye-tracking systems provide a more granular measure of attention shifts and mind-wandering, but are often costly, complex, and prone to calibration and accuracy issues \cite{Hutt2017-vm, Hutt2021-mw}.

More direct physiological measures include heart-rate variability (HRV) \cite{Byrne1996-oy}, skin conductance (EDA) \cite{Boucsein2007-tr}, and electroencephalography (EEG) \cite{Pope1995-td, Kosmyna2019-xi, Yuksel2016-ae}. EEG is a non-invasive technique that records electrical activity generated by neuronal firing in the brain, typically via electrodes placed on the scalp. Among these methods, EEG stands out as the one that directly measures neural activity, providing real-time insights into alertness, attention, and cognitive workload in both controlled and real-world settings \cite{Berka2007-ly, Frey2018-bp}. Since learning is fundamentally a neurological process, EEG offers a unique advantage by capturing dynamic brain responses during information processing, making it particularly well-suited for assessing engagement beyond behavioral proxies. At the same time, EEG also has well-known limitations, including susceptibility to movement artifacts, limited spatial resolution due to volume conduction, and practical constraints related to the number and placement of scalp electrodes, which can affect signal quality and user comfort. These limitations are particularly relevant in educational settings, where prolonged use, comfort, and sensitivity to subtle movements can impact data quality and learner experience.

Engagement can be measured using EEG through oscillatory activity (frequency-based markers) and event-related potentials (ERPs). Frequency-based markers provide continuous insights into attention and cognitive workload, with alpha power (8–12 Hz) linked to relaxation and disengagement \cite{Grammer2021-jy, Grassini2022-cn}, beta power (13–30 Hz) associated with sustained attention and active problem-solving \cite{Pope1995-td}, and theta power (4–8 Hz) indicative of fatigue or reduced vigilance \cite{Fairclough2009-tq}. A widely used composite metric is the Engagement Index, defined as Beta / (Alpha + Theta), where higher values indicate greater attentional focus and cognitive engagement \cite{Apicella2022-so, Hassib2017-iu, Kosmyna2019-xi, Pope1995-td, Eldenfria2019-er, Khedher2019-si}. Alpha asymmetry reflects differences in alpha power between the two brain hemispheres and is often associated with approach motivation and active engagement \cite{Fairclough2009-tq, Vujic2023-yf}. Other markers include the Cognitive Load Index (Theta Fz / Alpha Pz) \cite{Byrne1996-oy, Holm2009-vx} and alpha peak frequency \cite{Parsons2021-js}, which have both been explored as indicators of cognitive effort and attentional processing efficiency. ERPs, in contrast, offer time-locked neural responses to stimuli, with key components such as P300 (reflecting attentional allocation and task relevance), N200 (linked to conflict detection and executive control), and error-related negativity (ERN) (indicating engagement in performance monitoring) \cite{Bechtereva1968-mz, Patel2005-fi, Warchol2023-jx}. 

\subsubsection{EEG-Based Engagement in Learning}
EEG-based engagement metrics have been applied in various educational contexts, from providing feedback to presenters \cite{Hassib2017-iu} to tracking cognitive effort in classroom and workplace environments \cite{Grammer2021-jy, Hassib2017-iw}. EEG has been shown to capture distinct patterns of student attention that differ from self-reports and teacher observations, offering a more objective measure of engagement across instructional activities \cite{Grammer2021-jy}. Studies have also linked higher engagement to better learning performance. For example, EEG monitoring during video lectures revealed significant fluctuations in attentional focus, suggesting that lecture design should account for these variations \cite{Davidesco2023-uh}. Similarly, in a reasoning task with medical students, engagement correlated with task performance, though the highest engagement was observed in students who struggled, likely reflecting heightened cognitive effort despite insufficient expertise. This aligns with Vygotsky’s Zone of Proximal Development, suggesting that while students were highly engaged, the task was beyond their current skill level, leading to cognitive overload rather than effective learning \cite{Khedher2019-si}.

EEG has also been explored for cognitive workload classification, with implications for future learning technologies. Andreessen et al. \cite{Andreessen2021-cw} trained an EEG-based model to distinguish high and low mental workload, suggesting its potential for adaptive learning systems that adjust reading materials based on cognitive load. Similarly, Apicella et al. \cite{Apicella2022-so} demonstrated a low-channel, wearable EEG system for detecting engagement, proposing its use as an input channel for adaptive teaching platforms. While these studies focus on monitoring engagement rather than adapting learning in real time, they lay the groundwork for neuroadaptive systems that dynamically adjust instruction based on cognitive states. The next section explores how such systems leverage EEG engagement data to personalize learning experiences.

\subsection{Neuroadaptive Learning Systems}

Neuroadaptive systems leverage real-time neurophysiological data, particularly from electroencephalography (EEG), to dynamically adjust instructional content or interaction modalities based on a learner's cognitive and emotional states. These closed-loop systems aim to optimize learning outcomes by continuously monitoring engagement and adapting pedagogical strategies accordingly.

Early approaches to neuroadaptive learning focused on adapting presentation styles based on user engagement. For instance, Pay Attention! \cite{Szafir2012-gh} employed an embodied storytelling agent that adjusted its voice volume and gestures in real time to recapture students’ attention when EEG signals indicated a drop in engagement. This approach significantly enhanced the recall performance of students, demonstrating the potential of adaptive presentation to influence learning outcomes. Similarly, EngageMeter \cite{Hassib2017-iu} provided real-time feedback to keynote presenters about their audience’s engagement levels, enabling dynamic adjustments in delivery style. However, while effective in maintaining attention, these systems were limited to modifying delivery methods without altering the learning content itself. Thinking Cap \cite{Mills2017-au} extends these ideas into an Intelligent Tutoring System (ITS) featuring an animated tutor agent that dynamically adjusts the complexity of its instructional dialogue with the student based on EEG-derived cognitive load measures. This approach ensures that learners are neither underwhelmed nor overwhelmed. The authors pre-scripted easy and difficult versions of the instructional content by altering text complexity dimensions such as narrativity, syntactic ease, and referential cohesion. The more recent Online Continuous Adaptation Mechanism (OCAM) \cite{Eldenfria2019-er} builds on these principles by continuously monitoring not just engagement but also concentration, cognitive load, and emotional arousal to dynamically adjust the content difficulty, pacing, and presentation style. This system has been shown to significantly increase learner concentration and engagement, highlighting the value of multi-dimensional cognitive measures in adaptive learning.

Beyond academic learning environments, Learning Piano with BACh \cite{Yuksel2016-ae} dynamically adapts the difficulty of piano exercises based on cognitive workload (measured via functional near-infrared spectroscopy), guiding learners into their zone of proximal development to optimize skill acquisition. Closed-loop systems have also been found effective for enhancing learning in perceptual-cognitive tasks, as demonstrated by Parsons et al. \cite{Parsons2021-js}, who improved performance by manipulating a 3D multiple object tracking (3D-MOT) task through real-time neurofeedback.

Other systems focused on providing real-time biofeedback to help users self-regulate their engagement and attention. AttentivU \cite{Kosmyna2019-xi}, for instance, combines an EEG headband with haptic feedback devices that vibrate subtly when engagement levels drop, effectively redirecting attention in both online and in-person learning contexts. Unlike content-adaptive systems, these approaches rely on external cues to prompt re-engagement rather than altering the learning material itself. Similarly, Joie \cite{Vujic2023-yf} introduces a joy-based brain-computer interface (BCI) that uses prefrontal alpha asymmetry—an EEG marker linked to positive emotional states—to control an endless runner game. By training users to consciously modulate their brain activity through strategies like imagining joyful scenarios, Joie highlights the potential of neuroadaptive systems to foster affective engagement alongside cognitive performance.

Across these systems, a shared limitation is that all content-driven systems rely on pre-scripted content that needs to be prepared by the researchers to allow for adaptation. NeuroChat overcomes this barrier by integrating generative AI, which can create new content adapted to the complexity and presentation style of the reader’s cognitive state and specific questions on the fly.

\subsection{Generative AI-BCI Systems}

The integration of generative AI with brain-computer interfaces (BCIs) is an emerging research area. While machine learning has long been used to analyze EEG data, most AI-enhanced BCI systems have focused on brain state classification rather than interactive, real-time adaptive applications. The introduction of generative AI expands the possibilities of BCIs beyond passive decoding, enabling dynamic content modulation and interactive adaptation. Early investigations propose that integrating LLMs with BCIs could significantly enhance human-computer interaction, benefiting both individuals with neurological conditions and healthy users \cite{Caria2024-ff}.

\subsubsection{Using Generative AI to Analyze EEG}

A major focus in AI-BCI research has been EEG-based brain decoding, where generative AI and machine learning models are used to encode and decode the neural signals underlying visual or auditory information processing \cite{Bai2025-bg, Guenther2024-wt, Wang2024-eq}. While these methods advance neural signal processing, they remain limited in real-time user interaction. Readers interested in these approaches can refer to a comprehensive review by Sabharwal and Rama (2024) \cite{Sabharwal2024-iw}. Beyond decoding, LLMs have been increasingly applied to EEG for brain state classification and assistive communication \cite{Zhang2020-dm}. In clinical applications, language model-enhanced BCI communication systems have significantly improved typing accuracy for ALS patients by up to 84\% in online BCI spelling sessions \cite{Speier2016-ow}. Subsequent approaches have demonstrated that LLMs can classify brain states at the word level from EEG data during reading tasks \cite{Zhang2024-zf, Hollenstein2018-sr}. 

Recent research has extended these applications to foundation models that generalize across EEG tasks. NeuroLM \cite{Jiang2024-dn} and Neuro-GPT \cite{Cui2024-xe} function as foundation models, pre-trained on large EEG datasets using self-supervised learning and task-based fine-tuning to develop multi-purpose EEG processing models. NeuroLM, trained on over 25,000 hours of EEG recordings (predominantly the clinical TUH corpus), aligns brain signals with text-based representations, enabling multi-task analysis in cognitive workload detection, emotion recognition, and sleep staging. Neuro-GPT, pre-trained solely on the TUH EEG corpus, applies GPT-style tokenization to EEG data, improving feature extraction and adaptability to small datasets. In contrast, EEG-GPT \cite{Kim2024-ll} and Lee \& Chung (2024) \cite{Lee2024-rr} focus on task-specific applications—EEG-GPT applies few-shot learning for EEG-based brain state classification, while Lee \& Chung fine-tune GPT-3.5 Turbo for intracranial EEG (iEEG) interpretation, mapping neural signals to cognitive states. Because all four models are trained predominantly (or exclusively) on clinical recordings, they may underperform on EEG from neurologically healthy or other non‑clinical participants, highlighting the need for cross‑domain evaluation or adaptation when deploying these models.

Other approaches have explored personal health and well-being. For instance, Sano et al. (2024) \cite{Sano2024-op} used LLMs to interpret EEG signals for sleep quality assessment, providing tailored recommendations. Similarly, EEG Emotion Copilot \cite{Chen2024-xt} integrates EEG with a lightweight (0.5B parameter) LLM to analyze EEG signals, identify emotional states, and generate automated clinical insights. \cite{Hu2024-oz} propose MultiEEG-GPT, a model that integrates EEG with multimodal data—such as facial expressions and audio—to enhance mental health assessments using LLM-based classification.

Additionally, generative AI techniques have been leveraged for data augmentation to enhance EEG-based model training \cite{Eldawlatly2024-et, Zhou2023-cf}. However, while these approaches highlight generative AI’s ability to process EEG data for individual adaptation, they focus on recognizing states and have yet to support real-time user interaction. 

\subsubsection{Artistic Applications Using EEG to Modulate Generative AI Outputs}

While most research has focused on analyzing EEG data, a growing field explores EEG as a control mechanism for real-time generative AI adaptation. Early explorations have emerged in artistic and creative applications, where EEG signals influence AI-generated media production.
For example, Imagination Engine \cite{Unknown2023-le} translates EEG activity into abstract visual art, while Real-Time Neuro-Augmented Cinema \cite{Bellemare-Pepin2024-ni} enables cinematic modifications via neurofeedback. Similarly, the Bio-Mechanical Poet \cite{Tholke2024-oh} maps real-time EEG signals to symbolic representations, creating immersive poetic audiovisual experiences. These projects demonstrate EEG’s potential to actively modulate generative outputs rather than merely classifying brain states.
However, in these early stages, the potential to use real-time neurofeedback to shape AI-generated textual interactions—particularly in education—remains largely unexplored.

\subsubsection{Neuroadaptive Generative AI Systems}

The most advanced neuroadaptive system integrating generative AI for real-time adaptation is AdaptiveCoPilot \cite{Wen2025-yk}, designed for expert pilots in virtual reality. AdaptiveCoPilot continuously adjusts visual, auditory, and textual cues based on real-time cognitive load assessments, optimizing performance in high-stakes environments. However, it relies on functional near-infrared spectroscopy (fNIRS) rather than EEG and is tailored for high-performance cognitive tasks rather than learning applications.

Despite rapid advancements in AI-enhanced BCI research, no existing system has leveraged EEG data to dynamically modulate LLM-driven chatbot interactions. This gap underscores the novelty of NeuroChat, one of the first systems to integrate EEG-based cognitive state tracking with generative AI in real-time.

\section{System Design}

We set out four core design goals to ensure that NeuroChat is an accessible, responsive, and effective neuroadaptive learning system:
\begin{enumerate}
    \item \textbf{D1: Wearable, Non-Invasive Brain Sensing}: NeuroChat employs a consumer-grade, non-invasive EEG headband to measure engagement in real-time. We opted for the 4-channel, consumer-grade Muse EEG headband by InteraXon [60], balancing signal reliability with ease of use. This design ensures that users can engage with the system without complex electrode setups or invasive procedures.
    \item \textbf{D2: Real-Time Adaptive Personalization}: To maximize learning effectiveness, NeuroChat provides continuous neurofeedback, dynamically tailoring chatbot responses based on current engagement levels. The system processes EEG data in real time to adjust content complexity, response style, and pacing when prompted, ensuring an adaptive and personalized learning experience.
    \item \textbf{D3: Seamless EEG Integration in Natural User Environments}: Given that most users interact with NeuroChat on a laptop or desktop computer, the system accommodates a stationary, movement-minimizing environment, which is ideal for EEG signal acquisition. This design choice minimizes motion artifacts, resulting in more reliable neurofeedback processing.
    \item \textbf{D4: Web-Based Accessibility and Low-Cost Implementation}: NeuroChat is designed to be fully browser-based, eliminating the need for complex server-side infrastructure and enabling plug-and-play usability. Users can access the system on any platform with minimal setup, making it scalable and accessible to a broad audience.
\end{enumerate}

\subsection{Interaction Flow}

NeuroChat is built as a web app that integrates real-time EEG data with an LLM chatbot to create adaptive, personalized responses. The user flow is as follows (Figure~\ref{fig:user-flow}):

\begin{figure*}[htbp]
    \centering
    \includegraphics[width=0.97\textwidth]{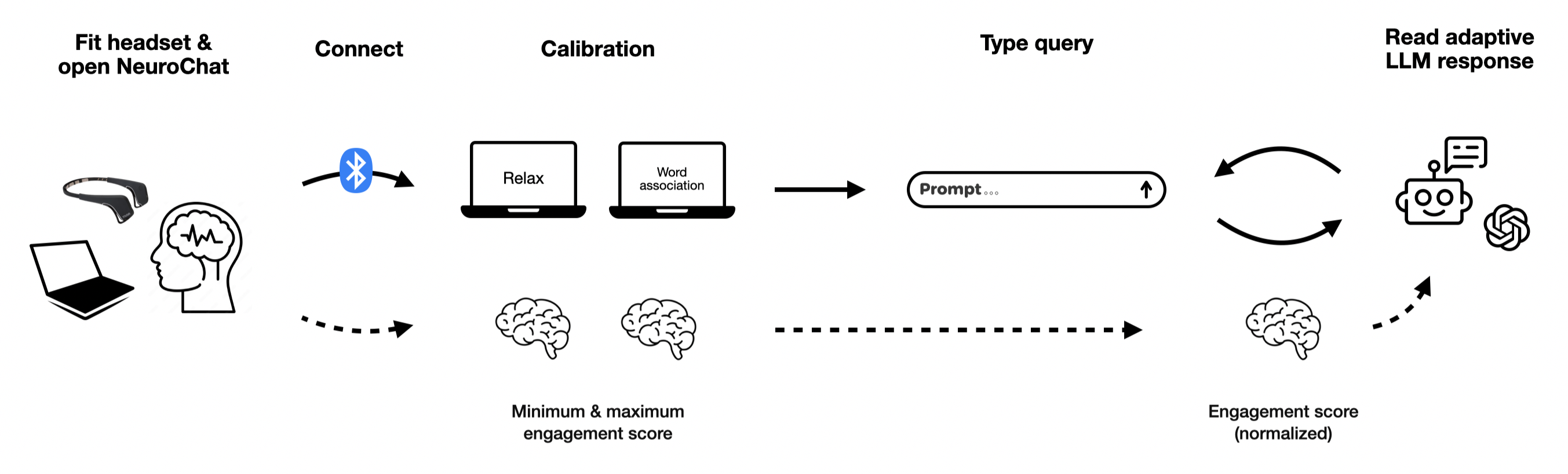}
    \caption{Overview of the NeuroChat system and user flow. The user connects the Muse headband, undergoes calibration, and interacts with the neurofeedback-driven LLM. Engagement scores are computed and inserted into the prompt unnoticed by the user.}
    \Description{The NeuroChat workflow in five steps using icons and arrows. The workflow begins with the user fitting the Muse EEG headband and opening the NeuroChat system, represented by a headband, laptop, and brain icon. The headband then connects via Bluetooth, followed by a calibration phase, shown with two laptop screens labeled "Relax" and "Word association" alongside brain icons indicating engagement levels. Next, the user types a query, represented by a text box labeled "Prompt...". Finally, the system provides an adaptive LLM response, illustrated by a chatbot and brain icon, showing that the response adjusts based on engagement. Arrows visually connect each step, depicting the process flow.}
    \label{fig:user-flow}
\end{figure*}

\begin{enumerate}
    \item \textbf{Connection:} The user fits the Muse 2 EEG headband on their head, ensuring skin contact. The NeuroChat web app searches for nearby Bluetooth devices via Web Bluetooth, establishes a connection to the Muse, and starts receiving data.
    \item \textbf{Calibration:} The user completes a 2-minute relaxation task to determine the engagement minimum (E\_min) and a 2-minute word association task for the engagement maximum (E\_max). These values are stored in the browser’s session storage for normalization.
    \item \textbf{Interaction:} Every second, the system computes an updated engagement score using a sliding window that averages the normalized engagement over the preceding 15 seconds. This window duration was selected to capture the user's sustained engagement while reading the chatbot’s most recent response, rather than capturing momentary fluctuations. Prior work has also established a 15-second interval as the minimum window for smoothing out variability in engagement signals \cite{Szafir2012-gh}.

    As soon as the user begins typing a new query, the most recent engagement score - representing the previous 15 seconds of engagement while reading the LLM's response - is frozen and inserted into the new query sent to the chatbot. While the user is typing, the engagement score is not updated, as our primary interest lies in the user's cognitive response to the LLM output, and because movement artifacts during typing would degrade EEG signal quality. These engagement scores are programmatically inserted but remain hidden from the user.

    \item \textbf{Interactive Response:} The user’s query, along with the embedded engagement score and a system prompt defining how to interpret it (Appendix \ref{app:neurochat-system-prompt}), is sent to the LLM provider, which returns a response dynamically tailored to the user’s cognitive state.

\end{enumerate}

\subsection{EEG Signal Processing}
\subsubsection{Device}

\begin{figure*}[htbp!]
    \centering
    \includegraphics[width=0.75\textwidth]{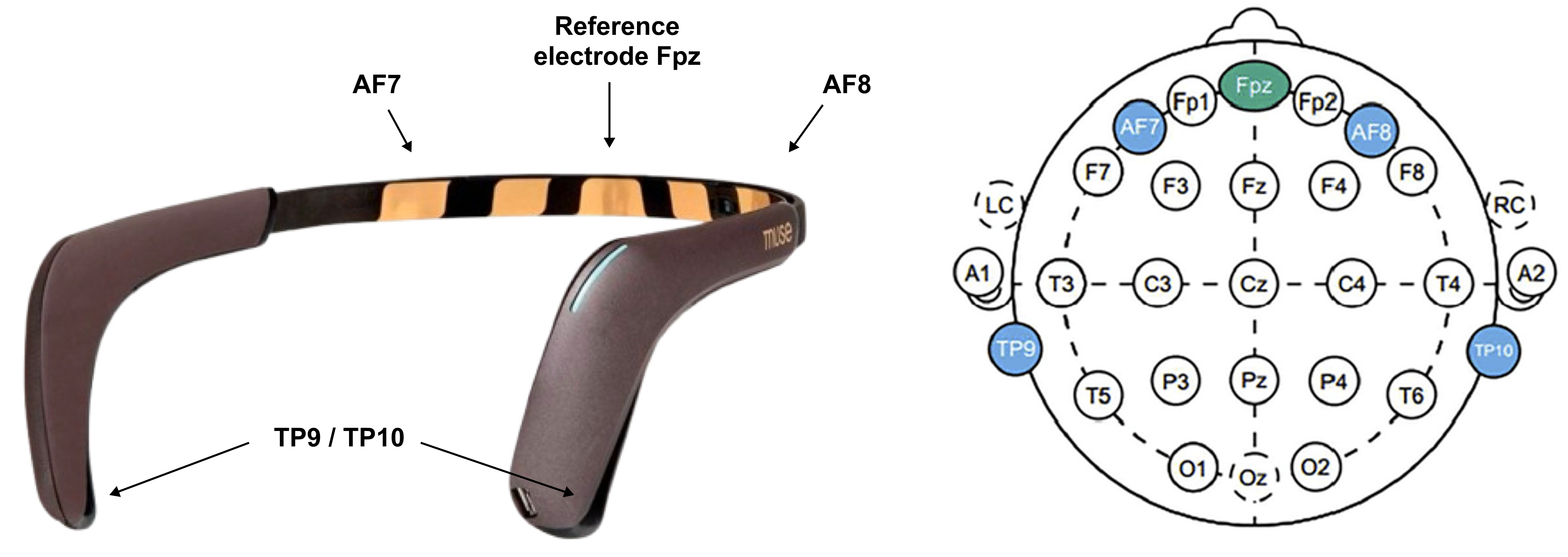}
    \caption{The Muse 2 EEG system made by InteraXon Inc (left). Electrode locations of Muse 2 headband according to 10-20 System (right); CC \cite{Kasperiuniene2016-oh}).}
    \Description{The Muse 2 EEG headband (left) with labeled electrode sites (AF7, AF8, TP9, TP10, and reference electrodes). On the right, a diagram maps these electrodes onto a standard 10-20 system, illustrating their placement on the scalp.}
    \label{fig:muse-system}
\end{figure*}

Our system uses the Muse 2 EEG headband, building on prior research that has leveraged consumer-grade devices with 1 to 6 channels to assess cognitive engagement in learning contexts (e.g., \cite{Hassib2017-iu, Kosmyna2019-xi, Szafir2012-gh, Vujic2023-yf}). The Muse 2 samples at 256 Hz and includes electrodes at Fpz, AF7, AF8, TP9, and TP10, following the 10-20 System (Figure~\ref{fig:muse-system}) \cite{InteraXon2025-el}. The Fpz electrode serves as the reference. EEG data is streamed via Bluetooth from the Muse device to a web app, which handles the connection using the open-source MuseJS library \cite{Shaked2021-zy}.

EEG systems vary along several key dimensions: the number and placement of electrodes (spatial resolution), the sampling rate (temporal resolution), and the type of electrodes (wet vs. dry, gel vs. fabric). While high-density research-grade systems offer superior signal fidelity and coverage, they are often expensive, obtrusive, and time-consuming to set up. In contrast, devices like the Muse 2 use a small number of dry electrodes to provide rapid, non-invasive EEG acquisition in real-world settings. Although Muse lacks the spatial resolution and noise rejection of clinical systems, it offers sufficient signal quality for measuring engagement-related frequency bands and frontal asymmetry, making it increasingly popular for field-based applications in human-computer interaction and learning research. Prior studies have validated the Muse platform against research-grade systems for detecting frequency-based indices of attention and workload \cite{Hassib2017-iu, Kosmyna2019-xi, Vujic2023-yf}. These capabilities make it a pragmatic and evidence-backed choice for a proof-of-concept, real-time neuroadaptive system like NeuroChat, where usability and deployment scalability are essential.

\subsubsection{Preprocessing}
The EEG data processing pipeline follows established methods from Hassib et al. \cite{Hassib2017-iu}, Kosmyna and Maes \cite{Kosmyna2019-xi},  Szafir and Mutlu \cite{Szafir2012-gh} and others. A bandpass filter (1–30 Hz) is applied to retain relevant neural activity while minimizing noise, and a 60 Hz notch filter removes power line interference. The data is then segmented into 1-second epochs with 250 ms intervals to enable continuous analysis with sufficient temporal resolution. Power spectral density is computed via fast Fourier transform (FFT), and band power is extracted for each frequency range to derive meaningful neural features.

\subsubsection{Engagement Score}
The engagement index (or engagement score) serves as the core metric of our system, enabling real-time quantification of cognitive engagement during mentally demanding tasks. First introduced by Pope et al. \cite{Pope1995-td}, this metric is computed as a ratio of key EEG frequency bands using the formula:

\begin{equation}
    E = \frac{\beta}{\alpha + \theta}
\end{equation}

where $\beta$ (11–20 Hz), $\alpha$ (7–11 Hz), and $\theta$ (4–7 Hz) correspond to EEG-derived neural oscillations. The index is based on the principle that higher beta power reflects heightened brain activity during cognitive tasks \cite{Boucsein2007-tr}. The beta frequency band is particularly associated with cognitive processes such as visual attention, motor planning, and active information processing, all of which indicate an engaged mental state. Conversely, increased alpha and theta activity typically correspond to lower cognitive engagement, with alpha waves linked to relaxation or passive states of rest \cite{Fairclough2009-tq, Grammer2021-jy, Grassini2022-cn}. 

The engagement index has been widely validated across various applications, including cognitive load assessments \cite{Freeman1999-hw}, visual processing studies, and sustained attention tasks \cite{Berka2007-ly}. It has also been applied in complex task environments such as the multi-attribute task battery (MATB) \cite{Pope1995-td}, which involves tasks like tracking, resource management, and communication. These studies demonstrate the metric's effectiveness in detecting attention shifts triggered by external stimuli \cite{Abdelrahman2015-pb, Dan2017-ir}.

Following our preprocessing pipeline, we extract frequency bands for each 1-second epoch and average them over a 15-second sliding window. Averaging over a time window allows us to assess a user’s engagement over a meaningful duration while they read and process the LLM’s output, rather than capturing momentary fluctuations. We selected a 15-second window that updates every one second following the established method by Szafir and Mutlu \cite{Szafir2012-gh} to account for variations in reading speed, ensuring sufficient time for users to engage with the response. Unlike previous studies, we opted against exponentially weighted moving averages, as our focus is on sustained engagement throughout a task rather than transient cognitive spikes.

Finally, we normalize the engagement score following the method proposed by Kosmyna and Maes \cite{Kosmyna2019-xi}. This process requires establishing a minimum and maximum engagement score for each participant, which we derive from a brief calibration phase conducted prior to the main experiment. During calibration, participants completed two tasks, each lasting 2 minutes, designed to elicit low and high levels of cognitive engagement. The first was a relaxation task in which participants remained still and minimized cognitive effort, providing a baseline for low engagement. The second was a mental word association task, where participants generated words based on the final letter of the previous one (e.g., “elephant” → “tiger”). This cognitively demanding activity has been shown to reliably induce elevated mental effort in healthy users for EEG experiments \cite{Kosmyna2023-oi}. The lowest and highest engagement scores recorded during these two tasks are taken as the normalization minimum \( E_{\text{min}} \) and maximum \( E_{\text{max}} \), respectively.

The calibration phase also differs from the main task in its use of a 10-second sliding window, striking a balance between the 5-second windows used in earlier studies \cite{Hassib2017-iu, Kosmyna2019-xi, Szafir2012-gh} and the 15-second window employed in our main experiment. Using these calibration bounds, we compute the normalized engagement score \( E_{\text{norm}} \) as:

\begin{equation}
    E_{\text{norm}} = \frac{E - E_{\text{min}}}{E_{\text{max}} - E_{\text{min}}}
\end{equation}

where \( E \) is the raw engagement score averaged over the past 15 seconds.

\subsection{LLM Adaptation}
The mechanism by which NeuroChat responds adaptively to the user’s cognitive state is by embedding their engagement score into each query submitted to a Large Language Model. We used OpenAI’s GPT-4-turbo model, the latest at the time of the study. Illustrative examples of how NeuroChat modulates its responses at different engagement levels are provided in Appendix~\ref{app:example-responses}.

To guide the model’s adaptations, we designed a system prompt that provides heuristics for tailoring responses based on the user’s real-time engagement score. This prompt was developed through iterative pilot testing and informed by learning science principles as well as guidance from OpenAI’s \textit{Teaching with AI} framework \cite{OpenAI2023-eh} (see Appendix~\ref{app:neurochat-system-prompt}). Early versions that asked the model to optimize for engagement often led to overly casual and upbeat responses, reflecting a misinterpretation of the term. Reframing the engagement score as a proxy for cognitive load produced more pedagogically appropriate output. The system prompt directs the LLM to adapt by modulating depth, specificity, response format (e.g., long-form vs. bullet points), and conversational style (e.g., Socratic questioning, analogies, examples). These strategies were drawn from established educational methods, such as constructivism and deeper learning, aiming to promote active processing and learner-centered dialogue. While the prompt can still be refined, our approach represents a first step toward embedding learning science into LLM-driven neuroadaptive tutoring.

\begin{figure*}[htbp]
  \centering

  \begin{subfigure}[t]{\linewidth}
    \centering
    \includegraphics[width=0.8\textwidth]{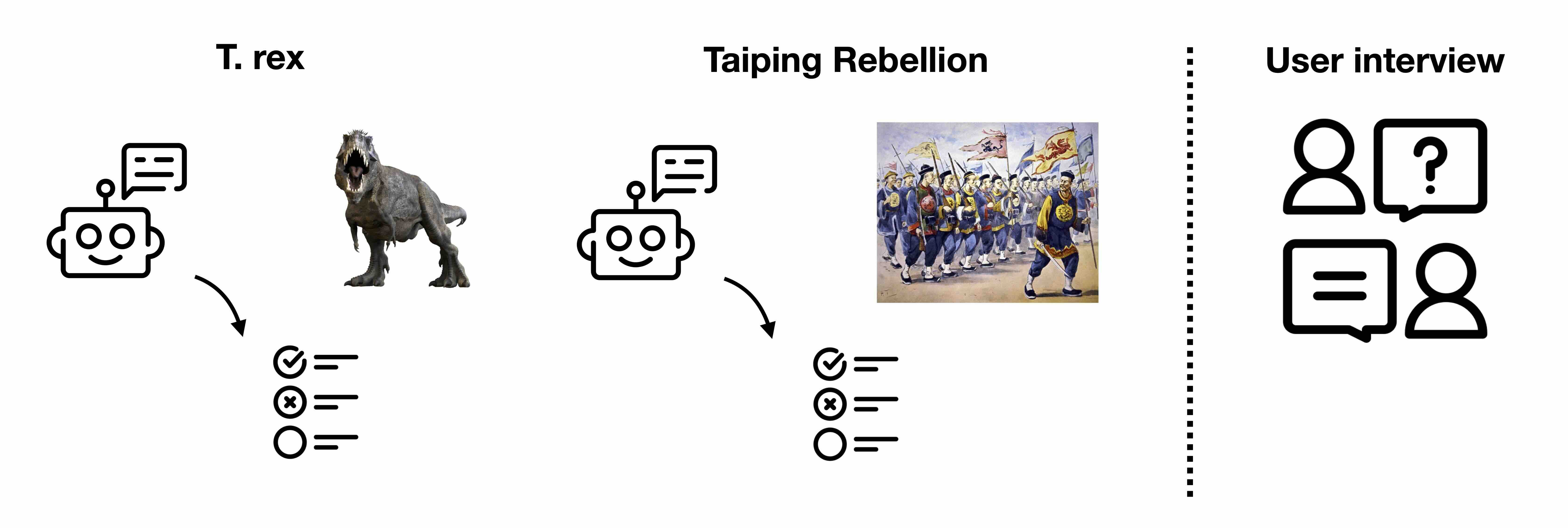}
    \caption{Learning task structure and experimental conditions.}
    \label{fig:study-tasks}
  \end{subfigure}

  \vspace{0.5em}

  \begin{subfigure}[t]{\linewidth}
    \centering
    \includegraphics[width=0.85\textwidth]{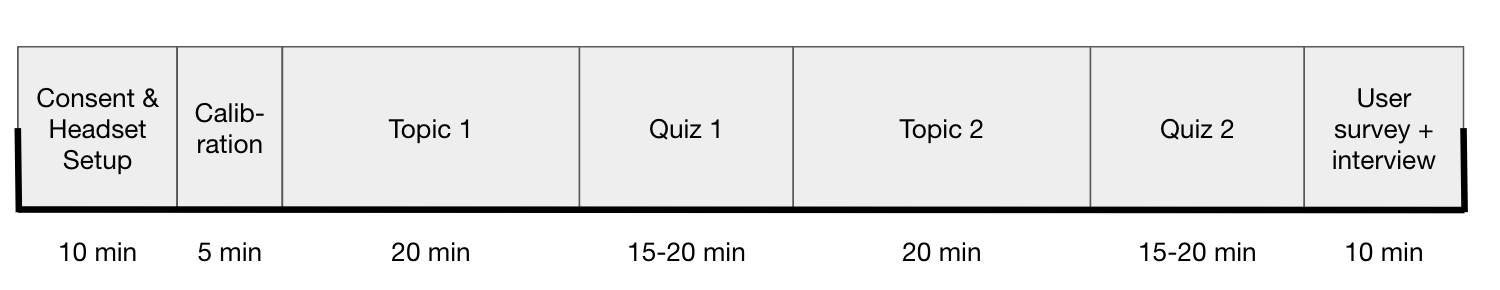}
    \caption{Study timeline and protocol.}
    \label{fig:study-protocol-flow}
  \end{subfigure}

  \caption{
    \textbf{Overview of study design.}
    (\textbf{a}) Visual representation of the experimental tasks. Participants interacted with two AI tutor conditions (NeuroChat and control) across two topics (T. rex and Taiping Rebellion), followed by topic-specific quizzes and a final user interview. 
    (\textbf{b}) Timeline of the full study session, including consent, EEG calibration, counterbalanced topic-condition blocks, quizzes, and concluding survey and interview.
  }
  \Description{
    (a) A diagram showing two learning topics (T. rex and Taiping Rebellion), each linked to an AI tutor interaction (either NeuroChat or control), followed by icons representing knowledge quizzes and a final interview. 
    (b) A horizontal timeline of the full study, with labeled blocks showing stages such as Consent & Headset Setup (10 min), Calibration (5 min), Topic 1 + Quiz 1, Topic 2 + Quiz 2, and User Survey & Interview (10 min).
  }
  \label{fig:study-protocol}
\end{figure*}

\subsection{User Interface}
The NeuroChat user interface (UI) consists of four key components (Figure \ref{fig:neurochat-ui}):

\begin{figure*}[htbp]
  \centering
  \includegraphics[width=0.8\textwidth]{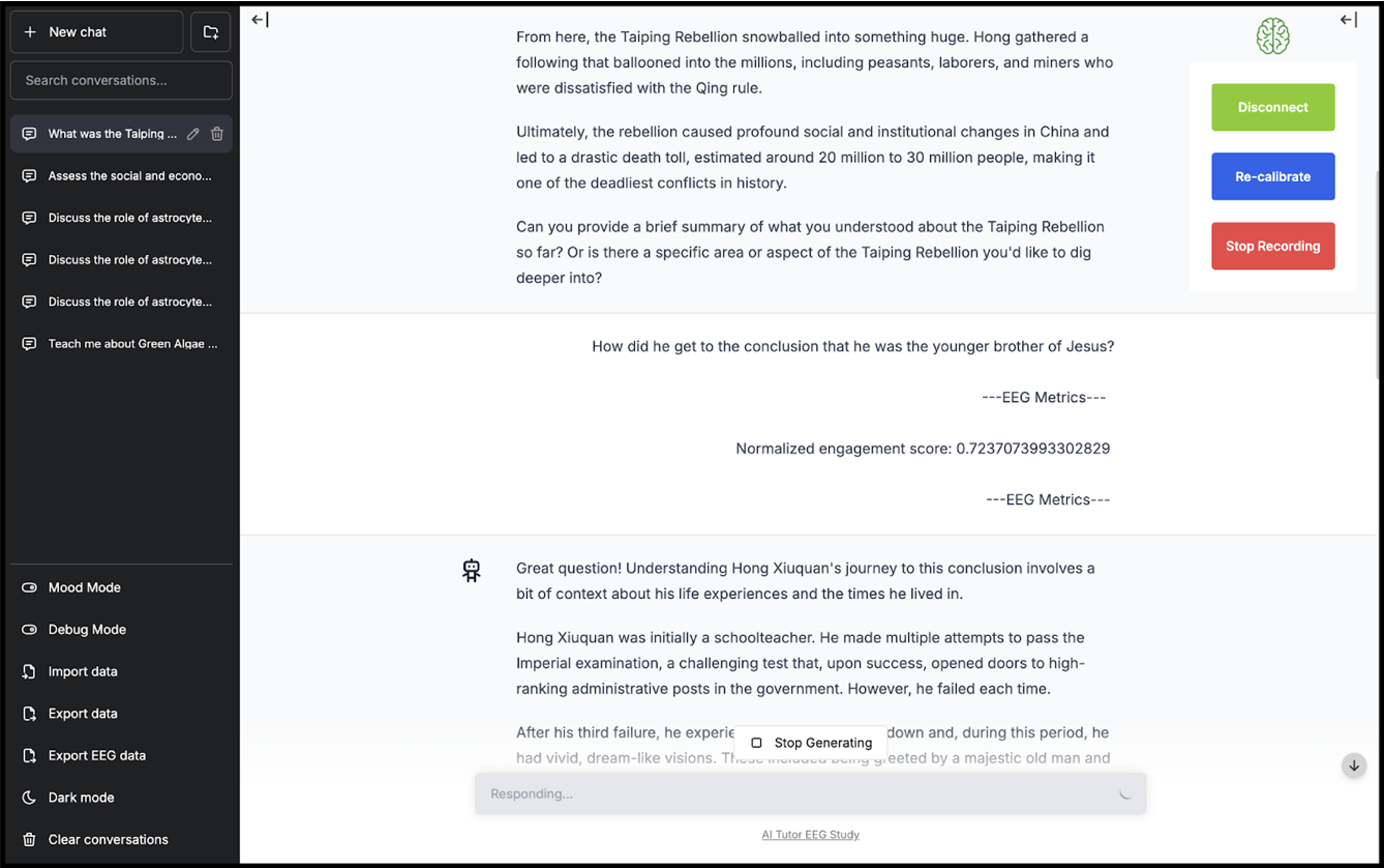}
  \caption{NeuroChat user interface with exposed EEG metrics in the user prompt and experimenter control menu. The connection to the Muse EEG device is made through the Brain Widget located in the top right corner. “Mood mode” activates the EEG metric injection into the user’s prompts, and turning off “Debug mode” allows the experimenter to hide these from the user. Chats, raw and filtered EEG data, and computed EEG metrics from the Muse device are stored in the browser’s native IndexedDB and can be exported from the Settings panel.}
  \Description{Screenshot of the NeuroChat UI. It features a brain widget with buttons to Connect, (Re-)Calibrate, and Start/Stop Recording; a main chatbot area, similar to ChatGPT; and a menu bar with past chats and settings.}
  \label{fig:neurochat-ui}
\end{figure*}

\begin{enumerate}
    \item \textbf{Brain Connect Widget} (Top Right Corner): Allows users to connect or disconnect the EEG device, calibrate or recalibrate the system, and start or stop EEG recording to compute the engagement index.
    \item \textbf{Calibration Modal} (Full Screen) – Appears only after the Muse headset is connected and provides instructions for a 2-minute relaxation phase followed by a 2-minute mental word association task. If the EEG connection is lost, users can restart or resume the completed relaxation phase.
    \item \textbf{Chatbot Interface} (Main Screen Area) – Functions similarly to ChatGPT, displaying an alternating conversation between the user and the AI tutor.
    \item \textbf{Menu Sidebar} (Left-Hand Side) – Contains chat history where experimenters can manage past LLM conversations by creating folders, renaming chat titles, and deleting individual chats. Also features the Settings panel, which provides options to toggle “Mood Mode” (enabling LLM adaptation), activate “Debug Mode” (revealing EEG metrics in the UI to the experimenter), import/export chat history, download EEG data (from the browser’s IndexedDB), switch between dark and light mode, and reset the chat history for a new user session. The terms “Mood Mode” and “Debug Mode” were intentionally chosen to provide visual cues to the experimenters while being vague enough to the participants, as they were not informed about the neuroadaptive functionality of the system.
\end{enumerate}

\section{Methodology}

\subsection{Hypotheses}
Based on prior research, we formulate the following hypotheses:
\begin{itemize}
    \item (H1) Objective Engagement: NeuroChat will elicit higher engagement levels than interaction with a standard GPT model, as measured by EEG-derived engagement scores.
    \item (H2) Subjective Engagement: Participants will report greater subjective engagement and satisfaction with NeuroChat, perceiving it as more engaging and effective than a traditional AI tutoring model.
    \item (H3) Learning Outcomes: Participants using NeuroChat will achieve higher scores on post-interaction learning assessments compared to those using the standard GPT model.
\end{itemize}

\subsection{Participants}
Thirty participants (15 female, 13 male, 2 non-binary or other), predominantly from academic backgrounds, were recruited for this study (M = 32.4 years, median = 30) and compensated with a \$50 Amazon gift card. The study received approval from MIT's institute’s ethical review board (protocol no. 21070000428).

\subsection{Study Design and Protocol}

We adopted a within-subject study design after pilot studies revealed significant individual differences in interactions with the AI chatbot. This design allowed each participant to serve as their own control, minimizing variability and enabling direct performance comparisons between the NeuroChat experimental condition and the control condition. The control condition consisted of a regular GPT chatbot, which was prompted to act within an AI tutoring task via its system prompt for fair comparison (see Appendix \ref{app:control-system-prompt}). Participants were not informed about the neuroadaptive nature of the experiment.

As study topics, we selected the \textit{Tyrannosaurus rex (T. rex)} and the \textit{Taiping Rebellion}. These topics were chosen to minimize prior topic bias while allowing room for facts and explorative interpretation. Although the T. rex is widely recognized, most people lack in-depth knowledge about the dinosaur. Similarly, despite its historical significance, the Taiping Rebellion is rarely emphasized in Western education. Both topics provided sufficient complexity and depth for meaningful engagement within the 20-minute learning session while still remaining accessible for participants. Condition and study topic order were counterbalanced using a Latin square design.

Before the session, participants signed a consent form and turned off their electronic devices. They were fitted with a Muse EEG headband, and signal quality was verified via the Muse EEG app \cite{InteraXon2024-iw}. Participants were instructed to minimize movement to reduce motion artifacts. 

The study lasted about 2 hours and proceeded as follows (Figure~\ref{fig:study-protocol}):

\begin{enumerate}
    \item \textit{Pre-Session Measures}: Participants completed a background questionnaire assessing their alertness and previous experience with AI chatbots. A brief EEG calibration phase followed (2 minutes relaxation, 2 minutes mental exercise).
    
    \item \textit{AI Chatbot Interaction}: Participants engaged with the chatbot for 20 minutes on their first assigned topic, with the goal of ``learning as much as possible.'' To guide exploration, they received starting pointers—e.g., characteristics, behavior, and archaeological research for T. rex and historical context, significance, and consequences for the Taiping Rebellion. Participants were free to focus on aspects they found interesting.
    
    \item \textit{Knowledge Assessment}: Immediately after the chatbot interaction, participants completed a quiz consisting of fill-in-the-blank and multiple-choice (MCQ) questions, followed by a 15-minute essay to assess understanding (see Appendix \ref{app:quiz-essay-questions}). To prevent preparatory bias, participants were not informed about the quiz beforehand. The same quiz was used across conditions.
    
    \item \textit{Break \& Condition Switch}: Participants took a short break before repeating the process with the second topic and condition. EEG data and chatbot interaction logs were continuously recorded.
    
    \item \textit{Final Survey \& Interview}: Participants completed a post-study user survey and a semi-structured interview focusing on their subjective engagement and experience across conditions. Interviews were thematically analyzed.
\end{enumerate}

\subsection{Evaluation}
Assessing learning outcomes requires a multifaceted approach. Factual recall was assessed through multiple-choice (MCQ) and fill-in-the-blank questions, covering information likely encountered during topic exploration. Meanwhile, creative synthesis was evaluated via mini-essays requiring recall, critical thinking, and analysis. Question complexity varied to reflect a range of difficulty levels appropriate for participants with minimal prior knowledge engaging for under 20 minutes. For the essay task, participants had 15 minutes to write and could choose from a set of prompts or create their own.

One author manually scored the quiz blind. Fill-in-the-blank questions were graded with 2 points each, with partial scores for semi-correct answers, and multiple choice questions (MCQ) were given 1 point per correct option. Since we only wanted to grade responses that had come up in the chat interaction, an automated keyword detection script scanned a participant’s message history for the presence or absence of a question, which was checked manually. Answers not covered were excluded from the participant’s total score, leaving us with proportional participant scores for comparison.

The 15-minute mini-essays were graded blind by a professional high school English teacher on a five-point scale across four categories: Content, Structure and Organization, Language and Style, and Accuracy (Spelling and Grammar). The Content category was given double weighting when calculating the final score.

In addition to objective assessments, participants completed a post-study user survey and were interviewed by one of the three authors in a semi-structured interview lasting up to five minutes. The interviewers took note of key quotes and sentiments and subsequently cross-read each other’s notes and discussed additional takeaways. The open-ended survey responses and interview notes were then compiled and analyzed using reflexive thematic analysis, following the methodology outlined by Braun \& Clarke \cite{Braun2012-ej}. This analysis was undertaken by the first author through multiple rounds of descriptive coding.

\section{Results}

To evaluate the effects of NeuroChat on engagement and learning outcomes, we conducted analyses on EEG engagement scores, learning assessments, and user feedback. Our results address three key areas: (1) cognitive engagement (EEG-derived engagement index), (2) user-reported engagement, and (3) learning performance (quiz and essay scores).

\subsection{Cognitive Engagement}

Since the system processes EEG data in real-time, engagement scores were computed dynamically during each session. Six participants were excluded due to missing or poor-quality EEG signals, and one additional participant was removed due to non-compliance with instructions. This resulted in 24 participants for analysis.

\begin{figure}[htbp]
  \centering
  \includegraphics[width=\linewidth]{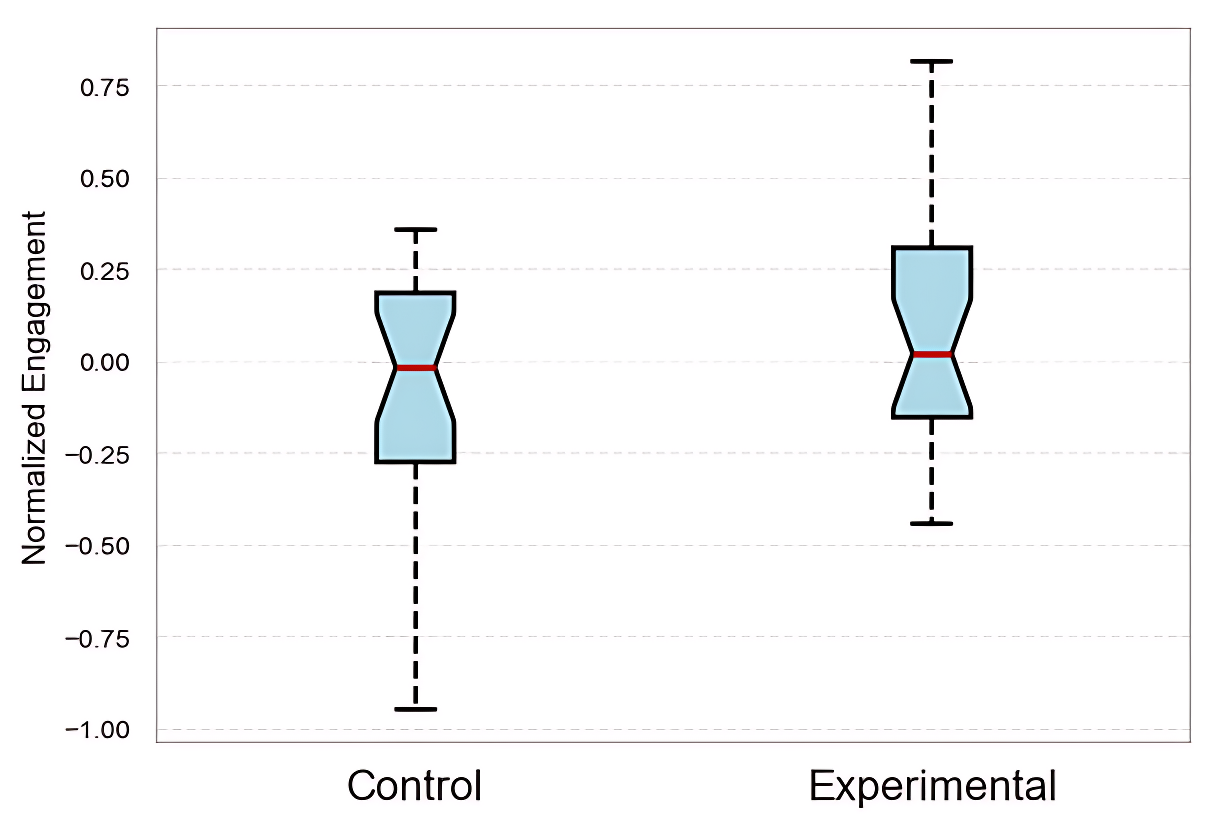}
  \caption{
    Overall distribution of normalized EEG engagement scores by condition. 
    Participants in the Experimental (NeuroChat) condition showed higher average engagement than in the Control condition. 
    Notches represent approximate 95\% confidence intervals around the medians.
  }
  \Description{
    Notched boxplot comparing normalized EEG engagement scores between Control and Experimental conditions. 
    The median for the Experimental condition is higher, and the distribution is slightly more positively skewed. 
    Each box represents the central 50\% of values, with whiskers extending to non-outlier range.
  }
  \label{fig:engagement-simple}
\end{figure}

For preprocessing, we removed missing values and extreme outliers (beyond 3× standard deviation) and manually inspected the engagement data, excluding segments with signal disconnections. Despite participant-level calibration, engagement scores exhibited relatively high between-subject variability (intra-class correlation coefficient (ICC) = 36.5\%, $p < 0.05$), which may reflect residual calibration inconsistencies or individual differences in EEG signal quality and responsiveness. To account for individual baseline differences while preserving within-subject variability, we applied z-score normalization, adjusting each participant’s engagement scores based on their mean and standard deviation across both conditions. This allowed for a direct comparison of relative engagement differences between NeuroChat and the control condition. Figure~\ref{fig:engagement-simple} provides a visual summary of normalized engagement scores across conditions, showing increased engagement in the experimental condition.

Given the repeated-measures design, where each participant contributed data across multiple conditions with occasional missing values, we fit a Linear Mixed Model (LMM) to examine the effect of experimental condition (Condition) on normalized engagement ($E_{\text{norm}}$). An initial model with Condition as the sole fixed effect yielded a marginally significant effect on engagement ($\beta = 0.186$, $p = 0.063$). 

Based on post-session interviews, many participants reported adjusting their approach in the second session once they became aware of the quiz component, regardless of condition. To account for this task adaptation, we added Order as an additional fixed effect. In the extended model, the effect of Condition reached statistical significance ($\beta = 0.216$, $p = 0.029$), while the effect of Order was marginal ($\beta = 0.181$, $p = 0.068$). The model coefficients are summarized in Table~\ref{tab:model-summary}. Although Order is not generalizable beyond this specific study design, we included it as a covariate to control for within-subject confounding in our counterbalanced experimental setup.

\begin{table*}[htbp]
  \centering
  \caption{Model summary for normalized engagement scores, examining the effects of Condition and Order using a Linear Mixed Model (LMM).}
  \begin{tabular}{lccccc}
    \toprule
    \textbf{Predictor} & \textbf{Coefficient ($\beta$)} & \textbf{Std. Error (SE)} & \textbf{z-value} & \textbf{p-value} & \textbf{95\% CI [Lower, Upper]} \\
    \midrule
    Intercept & -0.382 & 0.171 & -2.230 & 0.026 & [-0.718, 0.046] \\
    Condition (Experimental) & 0.216 & 0.099 & 2.185 & 0.029 & [0.022, 0.410] \\
    Order & 0.181 & 0.099 & 1.828 & 0.068 & [-0.013, 0.375] \\
    \bottomrule
  \end{tabular}
  \Description{
    Linear Mixed Model results, showing that the experimental condition significantly increases normalized engagement scores. 
    The Condition variable has a positive coefficient with $p = 0.029$, while Order has a marginal effect. 
    The intercept is negative and significant, reflecting the baseline level of z-scored engagement.
  }
  \label{tab:model-summary}
\end{table*}

As expected, the participant-level random effect variance was low ($\sigma^2 = 0.001$), reflecting the impact of z-score normalization on inter-individual differences. We also tested for effects of study topic, age, education level, chatbot experience, familiarity, usage frequency, and prompt engineering skill. None of these factors had a significant effect on engagement ($p > 0.1$).

\begin{figure*}[htbp]
  \centering
  \includegraphics[width=0.90\textwidth]{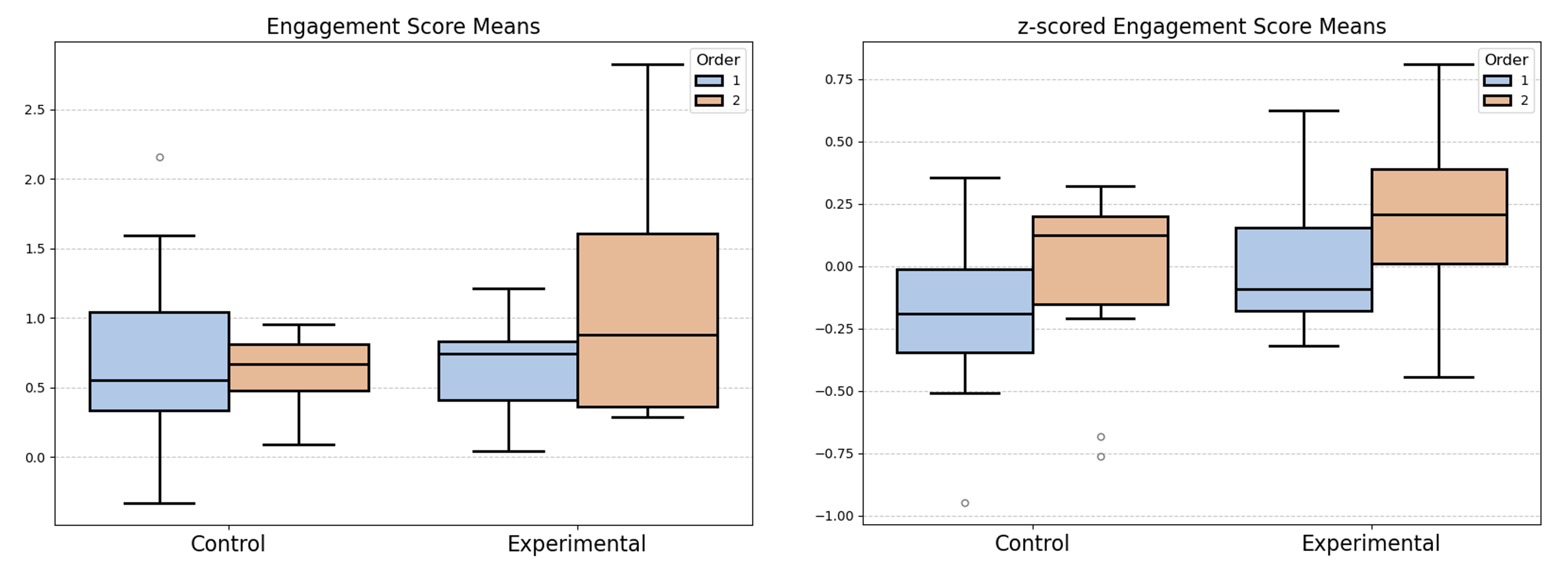}
  \caption{
    Distribution of engagement score means by condition and task order. 
    Left: Raw engagement score means. 
    Right: Z-scored engagement score means to enable comparability across participants.
  }
  \Description{
    Two boxplots comparing engagement scores across Control and Experimental conditions, separated by Order (1 = first session, 2 = second session).
    The left plot shows raw means; the right shows z-score normalized means. 
    In both, the Experimental condition, especially Order 2, shows greater central tendency and spread, suggesting increased engagement and variability.
  }
  \label{fig:e_score-means}
\end{figure*}

To further explore the effect of task order on engagement, we visualized participant-level engagement score means across conditions and orders. As shown in Figure~\ref{fig:e_score-means}, engagement was generally higher in the experimental condition, especially in the second session (Order 2), supporting the hypothesis that NeuroChat boosts sustained engagement over time.

In summary, when controlling for task order, participants in the NeuroChat condition exhibited significantly higher normalized engagement scores than in the control condition, indicating an association between neuroadaptive interaction and increased cognitive engagement.

\subsection{Learning Test Performance}

To assess whether NeuroChat improves learning outcomes (H3), we compared participants' quiz and essay performance. For the quiz, mean proportional scores showed no significant difference between conditions (E = 61.02\%, C = 60.66\%). Likewise, for the essay, participants in the experimental condition averaged 18.27 points, compared to 17.69 points in the control condition, indicating no notable difference between groups.

\subsection{Qualitative Reflections: Interviews and Surveys}

To explore how participants experienced the neuroadaptive system, we conducted semi-structured interviews with all 24 participants and performed a thematic analysis grounded in their words. Participants compared NeuroChat to a non-adaptive control chatbot and reflected on differences in tone, interaction style, usefulness, and engagement. Six central themes emerged, capturing a progression from surface-level impressions to deeper tensions in learning experience design. These findings provide supporting evidence for our hypothesis (H2) that participants using NeuroChat report higher levels of engagement and satisfaction than those using a non-adaptive chatbot.

\subsubsection{A More Human, More Enjoyable Chatbot}

Across participant reflections, 14 out of 24 described NeuroChat as more lifelike, creative, and engaging than the control chatbot. Participants used terms like “fun,” “enjoyable,” “fluid,” and “human-like” to describe the interaction. These reactions reflected not only linguistic tone but also perceived personality and presence. P30 characterized NeuroChat as “very fun, like a tour guide,” while P31 exclaimed, “Oh, I loved the second one! I really liked how it was saying how I was feeling.” P17 noted that “the interaction [with NeuroChat] felt more fluid and prone to fun facts, so more enjoyable.”

Even participants who ultimately preferred the control chatbot for clarity or factual density acknowledged that NeuroChat made the session more engaging. These positive reactions were consistent across many interviews: NeuroChat’s tone, curiosity-driven prompts, and personable voice made it feel like a conversational partner rather than just a tool.

\subsubsection{Different Minds, Different Modes}

Learners brought different expectations and preferences to the interaction, revealing contrasting cognitive styles and learning dispositions. Participants varied widely in how they evaluated the two systems, revealing meaningful differences in learning preferences and cognitive styles. While some found NeuroChat’s narrative and adaptive tone to be more readable and enjoyable, others preferred the control chatbot’s structured, information-dense format. These divergences were often tied to participants’ self-described learning dispositions.

Eight participants (P7, P10, P13, P14, P17, P19, P24, P26), especially those who identified with STEM or analytic learning styles, explicitly framed the control chatbot's more structured, "bland," "textbook-style" response style as more aligned with their preferred mode of learning. “I am a scientist, I like that [style of reading] more,” said P10. P13 similarly noted, “This type of knowledge and reading is what I am more used to and practice daily.” P19 described the control chatbot as “more concise,” “clear,” and “structured,” and P14 remarked that it was better “if you want to learn facts quickly.”

By contrast, 6 participants (P2, P14, P17, P27, P28, P31) expressed a preference for NeuroChat’s more accessible, conversational tone and found the control chatbot's response style "boring" and "difficult to consume." P28 called NeuroChat “more conversational and interesting,” while P14 appreciated that it was “more like an essay.” Similarly, while some participants described NeuroChat's responses as "basic," others found it to be "accessible" and "more readable." 

These differences underscore a broader point: learners bring their own expectations and preferences into AI-mediated education, and those preferences shape how they interpret the system’s usefulness and tone. Neuroadaptive and regular AI tutors, especially those intended for general audiences, may benefit from being aligned in style with learners' personal learning preferences, for instance, by offering presets or learning user preferences over time. 

\subsubsection{Tutor or Tool? Negotiating the Role of an Adaptive AI}

Participants varied in their interpretations of what NeuroChat was — some saw it as a tutor or companion, while others expected it to act like a neutral, responsive information source. This interpretive lens shaped their reactions to the system’s style and behavior.

Participants noted NeuroChat’s neuroadaptive tutor behavior: gently guiding the learner, asking follow-up questions, offering encouragement, and nudging users to think. For eleven participants (P1, P2, P10, P14, P19, P24, P26, P28, P30, P31, P33), NeuroChat felt “human-like,” “tutor-y,” or as if it had a personality. “It mimics a real person sort of feedback,” said P1. P19 described it as “more eager and tutor-y,” and P26 remarked that it “tried to act more like an elementary/middle school teacher.” P31 even speculated that the system was adapting in real time: “It was prompting ME... It sees that you’re not engaged and brings that to your attention and recommends ways to pursue the topic.” Participants who interacted with NeuroChat conversationally generally enjoyed the interaction more.

Yet this same behavior frustrated others, especially those who viewed the chatbot as a straightforward information retrieval tool. Those who approached NeuroChat as a Q\&A engine often found its tutoring behavior distracting, "annoying," and "redundant," especially when it asked follow-up questions instead of providing a direct answer. “I want to get information from it, I don’t want it to ask me questions,” said P14. P1 explained, “I knew what I wanted to ask... didn’t respond to the bot’s questions.” These participants expected control, not a system that turned the conversation around on them. 

The tensions here point to divergent epistemic expectations - some users seek a Socratic guide, others a personalized factual wiki. These divergent interpretations suggest that neuroadaptive systems must balance instructional autonomy with user mental models and expectations for control.

\subsubsection{Exploration vs. Extraction}

Participants consistently differentiated between the chatbots not just in style but in use case. Many positioned NeuroChat as a better fit for open-ended exploration and curiosity-driven dialogue, while the control chatbot was favored for studying, memorization, or fact retrieval. P6 summarized this distinction clearly: “The difference between [NeuroChat] and [the control chatbot] was like the difference between learning for personal interest and studying for an exam.” The interaction with NeuroChat was perceived as less information-dense and was compared to having a fun conversation versus "attending a serious lecture" in the control condition (P32). Many participants expressed that the control's responses feel similar to a "textbook" and more "concise," which makes them better suited for memorization. In contrast, P17 and P33 noted that memorizing facts was more difficult with NeuroChat, as its responses were "more scattered" and conversational.

At the same time, several participants (P2, P6, P12, P16, P19, P27, P28) emphasized how NeuroChat encouraged curiosity and deeper thinking. “I liked how [NeuroChat] allowed me to explore,” said P27. P28 added, “The more questions I asked, the more there was to learn.” This distinction maps a spectrum of AI tutor functions—between curiosity-led exploration and outcome-driven extraction—each suited to different learning goals.

\subsubsection{Engagement First, Learning Later?}

While NeuroChat consistently boosted engagement and enjoyment, some participants questioned whether that engagement translated into learning. Several noted that its responses, while engaging, were occasionally verbose, repetitive, or too general. “It tended to repeat itself… was more generic… giving too much info even when I asked a short question,” said P19. P30 added that it “always responds to my prompt,” but often with similar content that could feel redundant.

A few participants also described cognitive strain. “[NeuroChat] was more conversational, leading to a higher stress level to provide an answer,” said P7. P18, who preferred the control chatbot for clarity, still noted, “Despite my previous answer, the [NeuroChat] session was more engaging.” 

This theme reflects a broader educational trade-off: more stimulating systems may require more mental effort, which may not always optimize for recall. Yet the increase in subjective and cognitive engagement reported by many participants suggests that affective involvement may still be a valuable outcome in its own right. These findings suggest that while engagement may not always predict immediate learning gains, it could play a critical role in sustaining motivation or enabling deeper understanding over time.

\subsubsection{Imagining More: User-Driven Adaptation and Multimodality}
While most participants focused on their immediate experience with NeuroChat, a smaller subset envisioned ways the system could be expanded to better support engagement and personalization. Four participants (P2, P14, P24, P31) experimented with prompt engineering to modulate the system’s tone and format, using requests such as “write it like a poem” or “shorten this with notation.” These strategies reflect an emerging form of learner agency, where users attempt to shape the tutor’s voice or behavior to suit their preferences better.

Another group of 4 participants (P2, P10, P19, P30) expressed a desire for more visual or interactive features, such as real-time maps or diagrams. Others noted that images or timelines could reduce the cognitive burden of reading and make the tutor feel more dynamic. These suggestions position NeuroChat not only as a conversational partner but as a potential multimodal learning environment—one that blends language with rich visual scaffolding.

While a minority raised these ideas, such user-driven acts of prompting and imagining reveal an emerging form of co-adaptation — where learners not only respond to system behavior but actively shape it.
\section{Discussion}

This study examined whether NeuroChat, a neuroadaptive AI chatbot, enhances engagement and learning outcomes by adapting its responses based on real-time EEG feedback. Our results confirm that NeuroChat successfully increased cognitive (EEG-measured) and self-reported engagement, demonstrating its feasibility as a neuroadaptive tutoring system. However, no significant differences were found in learning performance, indicating challenges in translating engagement into measurable knowledge gains. Below, we discuss these findings in the broader context of adaptive learning, generative AI, and brain-computer interfaces (BCIs) before exploring key challenges and future directions.

\subsection{Key Findings: Engagement Gains, Learning Outcomes, and User Perception}

Our results show that NeuroChat was associated with significantly increased EEG-measured engagement compared to a non-adaptive chatbot (p = 0.029), suggesting that real-time neuroadaptive feedback can enhance sustained attention during open-ended tutoring sessions. These findings align with prior systems such as Pay Attention! \cite{Szafir2012-gh} and EngageMeter \cite{Hassib2017-iu}, which modulated engagement through adaptive content delivery. NeuroChat extends this paradigm by adapting not just pacing or difficulty, but also tone, complexity, and response strategy in the context of an ongoing dialogue, marking a shift towards generative, closed-loop neuroadaptive interfaces.

Subjective user feedback supported this effect: participants described NeuroChat’s responses as more human-like, responsive, and attuned to their needs. However, preferences varied, as some users appreciated exploratory dialogue, while others favored concise, fact-driven responses. This reinforces the importance of incorporating user-configurable interaction settings to better support diverse learning styles.

Despite these engagement gains, NeuroChat did not significantly improve quiz or essay scores compared to the control condition. Several factors may explain this gap. First, EEG-based engagement primarily reflects attentional state, not necessarily deeper learning or retention. Second, unlike structured adaptive systems such as BACh \cite{Yuksel2016-ae}, NeuroChat offered open-ended, user-led exploration, with limited scaffolding or explicit feedback. Third, our study measured learning outcomes after a single 20-minute session, which may be insufficient to observe downstream effects on knowledge acquisition.

Our findings resonate with recent work by Leong et al. \cite{Leong2024-hu}, who investigated AI-enabled personalization in a vocabulary learning task. While personalized content generated by LLMs increased learners’ motivation, perceived autonomy, and sense of competence, it did not yield measurable learning gains in the short term. These parallels suggest that generative AI may be particularly effective at enhancing affective and motivational dimensions of learning - factors known to influence long-term learning behaviors - even if short-term retention gains are not immediately evident. As Leong et al. also note, such benefits may compound over time through repeated exposure, sustained motivation, or deeper personalization. Neuroadaptive systems like NeuroChat may thus be most impactful when deployed in sustained, personalized learning contexts that allow motivation, strategy, and conceptual understanding to develop over time.

More broadly, our results raise important questions about what types of learning generative neuroadaptive systems can support and how such learning should be measured. While our study focused on retention, many participants’ essays simply restated facts rather than synthesizing ideas. This may reflect not only the recency of content exposure but also the limited potential for deep integration within a single session. Future work should consider longitudinal studies and alternative learning metrics, such as conceptual understanding, problem-solving ability, or creative insight, which may be more aligned with the types of cognitive engagement that neuroadaptive interfaces like NeuroChat can foster.

\subsection{Challenges and Limitations}
A key challenge of the NeuroChat design lies in the high variability of learner preferences and engagement patterns. Participants demonstrated diverse reactions to interaction styles: some thrived in open-ended, empathetic conversations, while others preferred straightforward, fact-focused responses. These differences complicate the mapping between engagement scores and LLM adaptations, particularly in the absence of explicit learner modeling. Although our current system modulates output based on engagement levels, it lacks awareness of individual preferences or goals. Future systems should incorporate user-configurable settings for interaction style, response format (e.g., bullet points versus narratives), and persistent personalization mechanisms such as memory-based user profiles (e.g., OpenAI’s memory feature \cite{OpenAI2024-zr}).

This challenge is further reflected in how engagement values are integrated into the system. While the LLM receives real-time engagement scores, there is no fixed rule set that defines how different values should influence the response. During pilot testing, we explored providing explicit adaptation strategies based on engagement score ranges in the system prompt; however, we found that learners varied too widely in what they considered helpful or engaging. We still observed these user-specific differences with our more open and adaptable LLM prompt, as some users preferred, for instance, a more factual style, and others liked a more personable and empathetic approach. As a result, we adopted a more flexible approach: engagement scores are paired with general behavioral guidance, and the LLM is expected to adapt dynamically over time through interaction. While this design supports open-ended personalization, it introduces variability in behavior and complicates the process of standardized evaluation. We view this as a necessary tradeoff in developing more ecologically valid and user-centered neuroadaptive systems.

A related concern is the ambiguity of engagement itself. While increased engagement is often interpreted as a positive signal, prior research shows that it can also reflect cognitive overload, confusion, or frustration \cite{Eldenfria2019-er}. Future neuroadaptive tutor designs must ensure users remain in their zone of proximal development rather than pushing them beyond their capabilities. Our current system does not distinguish between productive and unproductive forms of engagement, nor does it adapt based on the learner’s prior knowledge, misconceptions, or cognitive load. Mapping a score of 0.3 versus 0.7 to meaningful instructional strategies remains undetermined without a deeper understanding of the learner’s context. Future systems could benefit from integrating complementary signals, such as task performance, self-reports, or behavioral metrics, to guide adaptation more accurately. 

Another critical limitation lies in how learning was operationalized and assessed. While we used quiz and essay tasks to measure retention, this may not fully capture the types of learning NeuroChat is best equipped to support. Measuring learning is inherently complex, particularly within a single short session, and retention is only one dimension. Neuroadaptive tutors, such as NeuroChat, may be better suited to support conceptual understanding, critical thinking, or creativity. However, these constructs are more challenging to measure and often require longitudinal designs, domain-specific rubrics, or qualitative evaluation. Although we included retention measures out of curiosity, our results suggest that short-term engagement gains may not translate into immediate performance improvements, underscoring the need for richer metrics and longer-term studies.

The quality of EEG signals further constrains system performance. While consumer-grade devices like the Muse 2 offer portability and ease of use, they are limited by low spatial resolution, dry electrodes, and high sensitivity to motion and muscle artifacts—including eye blinks, facial tension, and head movement. These noise sources can distort engagement metrics and reduce the reliability of real-time inference. Although our use of frequency-domain features and a 15-second sliding window helps mitigate this variability, these techniques cannot fully eliminate artifact-related noise. More robust neuroadaptive systems may benefit from multimodal sensing, artifact rejection techniques, or sensor fusion.

In addition to technical limitations, privacy remains a critical consideration. EEG signals are inherently biometric and can be used to uniquely identify individuals \cite{Ruiz-Blondet2016-qj}. Although our prototype runs entirely client-side and does not store or transmit engagement data externally, any future deployment at scale would require stringent privacy-preserving mechanisms.

Finally, this study was limited by a brief interaction duration (20 minutes) and a modest sample size (n = 24). While these parameters are consistent with prior BCI-HCI studies, they limit the generalizability of our findings and preclude strong claims about long-term learning outcomes. We present this work not as a conclusive evaluation, but as a proof of concept that demonstrates the feasibility and design potential of neuroadaptive language interfaces.

\subsection{Implications for Neuroadaptive Learning and AI-Powered Tutoring}

Traditional neuroadaptive learning systems relied on pre-scripted content, where researchers manually assigned learning materials to high- or low-engagement conditions. NeuroChat overcomes this limitation by leveraging generative AI to create content dynamically, enabling real-time adaptation tailored to individual users. This is a fundamental shift in adaptive learning—moving from rule-based, pre-mapped content to generative, personalized tutoring. Most LLMs require users to explicitly communicate their needs (e.g., "make this simpler"). NeuroChat infers engagement implicitly through neural signals, reducing the need for manual prompt engineering. This opens possibilities for more seamless and accessible personalized tutoring experiences, especially for users who may not know how to articulate their needs.

NeuroChat has particular relevance for self-directed learners, who often struggle to maintain engagement in unstructured settings. In 2021, over 220 million students enrolled in MOOCs, yet average completion rates remain below 15\% \cite{Onah2014-vk, Unknown2023-ln}. A neuroadaptive AI tutor that responds to cognitive engagement in real time could help sustain motivation and prevent dropout in these autonomous learning environments. Beyond education, such a system may benefit knowledge workers, particularly those who struggle with focus and information retention. Prior works have proposed LLM-based BCI for aiding individuals with learning challenges, including ADHD \cite{Huang2014-ch, Parsons2021-js, Caria2024-ff}. These applications suggest that EEG-informed generative systems could extend beyond the classroom to broader domains of cognitive support.

A recent EEG study on LLM-assisted essay writing found that while LLMs offer convenience, they were associated with reduced neural engagement, weaker brain connectivity, and lower essay ownership and recall compared to traditional writing or search-based methods \cite{Nataliya2025-wa}. These results highlight the risk that passive reliance on LLMs may hinder deep cognitive processing. However, this also points to a unique opportunity for neuroadaptive systems like NeuroChat: by using real-time engagement signals, such systems could help prevent cognitive overreliance, actively guiding learners toward sustained effort and deeper processing. Rather than acting as passive information sources, neuroadaptive tutors could serve as proactive partners—dynamically nudging learners into their zone of proximal development and fostering more active, effortful learning.

Future designs may benefit from a hybrid approach: implicit neural feedback combined with explicit user preference signaling, enabling co-adaptation between learner and system.

\section{Conclusion}
This paper presented NeuroChat, a proof-of-concept neuroadaptive AI tutor that uses real-time EEG engagement signals to guide a generative language model in adapting its responses. Our study demonstrates that NeuroChat increases cognitive and perceived engagement but does not significantly improve short-term learning performance. These findings highlight both the promise and challenges of brain-in-the-loop learning systems. As neuroadaptive interfaces evolve, we recommend that future work focus on longer-term studies, deeper personalization, and learning metrics beyond retention to fully realize the potential of neuroadaptive tutoring systems.

\begin{acks}
We thank Treyden Chiaravalloti for his valuable piloting support and insightful feedback. We also appreciate Protyasha Nishat’s expertise in signal processing and Nathan Whitmore’s comments on study design. Luisa Heiss’s thorough grading of the essays was instrumental in evaluating participant test performance without bias.
This research was supported by the MIT Jameel World Education Lab (J-WEL) Education Innovation Grant.
\end{acks}

\bibliographystyle{ACM-Reference-Format}
\bibliography{bibliography}

@ARTICLE{Ruiz-Blondet2016-qj,
  title     = "{CEREBRE}: A novel method for very high accuracy event-related
               potential biometric identification",
  author    = "Ruiz-Blondet, Maria V and Jin, Zhanpeng and Laszlo, Sarah",
  journal   = "IEEE Trans. Inf. Forensics Secur.",
  publisher = "Institute of Electrical and Electronics Engineers (IEEE)",
  volume    =  11,
  pages     = "1618--1629",
  month     =  jul,
  year      =  2016,
  doi       = "10.1109/tifs.2016.2543524",
  language  = "en"
}

@INBOOK{Braun2012-ej,
  title     = {Thematic analysis},
  author    = {Braun, Virginia and Clarke, Victoria},
  booktitle = {APA handbook of research methods in psychology, Vol 2: Research
               designs: Quantitative, qualitative, neuropsychological, and
               biological},
  publisher = {American Psychological Association},
  location  = {Washington},
  pages     = {57--71},
  date      = {2012},
  doi       = {10.1037/13620-004},
  isbn      = {9781433810053},
  url       = {https://sites.psych.ualberta.ca/IClab/wordpress/wp-content/uploads/2022/02/Braun12APAHandbook.pdf}
}

@ARTICLE{Dubey2024-md,
  title        = {Redefining cognitive domains in the era of {ChatGPT}: A
                  comprehensive analysis of artificial intelligence's influence
                  and future implications},
  author       = {Dubey, Souvik and Ghosh, Ritwik and Dubey, Mahua Jana and
                  Chatterjee, Subhankar and Das, Shambaditya and Benito-León,
                  Julián},
  journaltitle = {Med. Res. Arch},
  publisher    = {Knowledge Enterprise Journals},
  volume       = {12},
  issue        = {6},
  pages        = {5383},
  date         = {2024-06-24},
  doi          = {10.18103/mra.v12i6.5383},
  issn         = {2375-1916,2375-1924},
  url          = {http://dx.doi.org/10.18103/mra.v12i6.5383},
  urldate      = {2025-05-08},
  language     = {en}
}

@INBOOK{Kasperiuniene2016-oh,
  title     = {Affective Engagement to Virtual and Live Lectures},
  author    = {Kasperiuniene, Judita and Jariwala, Meet and Vaskevicius,
               Egidijus and Satkauskas, Saulius},
  booktitle = {Communications in Computer and Information Science},
  publisher = {Springer International Publishing},
  location  = {Cham},
  pages     = {499--508},
  date      = {2016},
  doi       = {10.1007/978-3-319-46254-7_40},
  issn      = {1865-0937,1865-0929},
  isbn      = {9783319462530,9783319462547},
  series    = {Communications in computer and information science},
  url       = {http://dx.doi.org/10.1007/978-3-319-46254-7_40},
  urldate   = {2025-05-09},
  language  = {en}
}

@ARTICLE{Nataliya2025-wa,
  title        = {Your brain on {ChatGPT}: Accumulation of cognitive debt when
                  using an {AI} assistant for essay writing task},
  author       = {Nataliya, Kosmyna and Eugene, Hauptmann and Yuan, Ye Tong and
                  Jessica, Situ and Xian-Hao, Liao and Beresnitzky, Ashly Vivian
                  and Iris, Braunstein and Pattie, Maes},
  journaltitle = {arXiv [cs.AI]},
  date         = {2025-06-10},
  url          = {http://arxiv.org/abs/2506.08872},
  urldate      = {2025-06-12}
}

@ONLINE{InteraXon2024-iw,
  title     = {Muse: Meditation \& Sleep},
  author    = {{InteraXon}},
  booktitle = {App Store},
  date      = {2024},
  url       = {https://apps.apple.com/us/app/muse-meditation-sleep/id849841170},
  urldate   = {2024-09-13},
  language  = {en}
}

@ARTICLE{de-Jong2010-ff,
  title     = "Cognitive load theory, educational research, and instructional
               design: some food for thought",
  author    = "de Jong, Ton",
  journal   = "Instr. Sci.",
  publisher = "Springer Nature",
  volume    =  38,
  pages     = "105--134",
  month     =  mar,
  year      =  2010,
  doi       = "10.1007/s11251-009-9110-0",
  language  = "en"
}

@ARTICLE{Deng2025-uk,
  title     = "Does {ChatGPT} enhance student learning? A systematic review and
               meta-analysis of experimental studies",
  author    = "Deng, Ruiqi and Jiang, Maoli and Yu, Xinlu and Lu, Yuyan and Liu,
               Shasha",
  journal   = "Comput. Educ.",
  publisher = "Elsevier BV",
  volume    =  227,
  pages     =  105224,
  month     =  "1~" # apr,
  year      =  2025,
  doi       = "10.1016/j.compedu.2024.105224",
  language  = "en"
}

@ARTICLE{Heung2025-xu,
  title     = "How {ChatGPT} impacts student engagement from a systematic review
               and meta-analysis study",
  author    = "Heung, Yuk Mui Elly and Chiu, Thomas K F",
  journal   = "Computers and Education: Artificial Intelligence",
  publisher = "Elsevier BV",
  volume    =  8,
  pages     =  100361,
  month     =  "1~" # jun,
  year      =  2025,
  doi       = "10.1016/j.caeai.2025.100361",
  language  = "en"
}

@ARTICLE{Essel2024-oa,
  title     = "{ChatGPT} effects on cognitive skills of undergraduate students:
               Receiving instant responses from {AI}-based conversational large
               language models ({LLMs})",
  author    = "Essel, Harry Barton and Vlachopoulos, Dimitrios and Essuman,
               Albert Benjamin and Amankwa, John Opuni",
  journal   = "Computers and Education: Artificial Intelligence",
  publisher = "Elsevier BV",
  volume    =  6,
  pages     =  100198,
  month     =  "1~" # jun,
  year      =  2024,
  doi       = "10.1016/j.caeai.2023.100198",
  language  = "en"
}

@ARTICLE{Mulyani2025-sj,
  title     = "Transforming education: exploring the influence of generative
               {AI} on teaching performance",
  author    = "Mulyani, Heni and Istiaq, Mohamad Azim and Shauki, Elvia R and
               Kurniati, Fitrina and Arlinda, Hanifia",
  journal   = "Cogent Educ.",
  publisher = "Informa UK Limited",
  volume    =  12,
  month     =  "31~" # dec,
  year      =  2025,
  doi       = "10.1080/2331186x.2024.2448066",
  language  = "en"
}

@ARTICLE{Schultz2025-on,
  title     = "This School Will Have Artificial Intelligence Teach Kids (With
               Some Human Help)",
  author    = "Schultz, Brooke",
  journal   = "Education Week",
  publisher = "Education Week",
  month     =  "6~" # jan,
  year      =  2025,
  language  = "en"
}

@TECHREPORT{The-Open-Innovation-Team-and-Department-for-Education2024-cs,
  title       = "Generative {AI} in education - Educator and expert views",
  author      = "{The Open Innovation Team and Department for Education}",
  institution = "UK Department for Education",
  month       =  jan,
  year        =  2024
}

@MISC{David-Game-College2024-gh,
  title        = "{GCSE} {AI} Adaptive Learning Programme",
  author       = "{David Game College}",
  year         =  2024,
  howpublished = "\url{https://www.davidgamecollege.com/courses/courses-overview/item/102/gcse-ai-adaptive-learning-programme}",
  language     = "en"
}

@MISC{KhanAcademy2025-on,
  title        = "Meet Khanmigo: Khan Academy's {AI}-powered teaching assistant
                  \& tutor",
  author       = "{KhanAcademy}",
  year         =  2025,
  howpublished = "\url{https://www.khanmigo.ai/}",
  language     = "en"
}

@MISC{OpenAI2024-bj,
  title        = "Introducing {ChatGPT} Edu",
  author       = "{OpenAI}",
  month        =  "30~" # may,
  year         =  2024,
  howpublished = "\url{https://openai.com/index/introducing-chatgpt-edu/}",
  language     = "en"
}

@MISC{Wang2024-eq,
  title        = "{EEG2Video}: Towards Decoding Dynamic Visual Perception from
                  {EEG} Signals",
  author       = "Wang, Yansen and Wang, Zilong",
  month        =  "24~" # dec,
  year         =  2024,
  howpublished = "The Thirty-eighth Annual Conference on Neural Information
                  Processing Systems (NeurIPS 2024)",
  language     = "en"
}

@ARTICLE{Guenther2024-wt,
  title     = "Image classification and reconstruction from low-density {EEG}",
  author    = "Guenther, Sven and Kosmyna, Nataliya and Maes, Pattie",
  journal   = "Sci. Rep.",
  publisher = "Springer Science and Business Media LLC",
  volume    =  14,
  pages     =  16436,
  month     =  "16~" # jul,
  year      =  2024,
  doi       = "10.1038/s41598-024-66228-1",
  language  = "en"
}

@ARTICLE{Sabharwal2024-iw,
  title        = "Comprehensive review of {EEG}-to-output research: Decoding
                  neural signals into images, videos, and audio",
  author       = "Sabharwal, Yashvir and Rama, Balaji",
  journal      = "arXiv [cs.CV]",
  month        =  "27~" # dec,
  year         =  2024,
  primaryClass = "cs.CV"
}

@INCOLLECTION{Bai2025-bg,
  title     = "{DreamDiffusion}: High-quality {EEG}-to-image generation with
               temporal masked signal modeling and {CLIP} alignment",
  author    = "Bai, Yunpeng and Wang, Xintao and Cao, Yan-Pei and Ge, Yixiao and
               Yuan, Chun and Shan, Ying",
  booktitle = "Lecture Notes in Computer Science",
  publisher = "Springer Nature Switzerland",
  pages     = "472--488",
  series    = "Lecture notes in computer science",
  year      =  2025,
  doi       = "10.1007/978-3-031-72751-1\_27",
  isbn      = "9783031727504,9783031727511",
  language  = "en"
}

@ARTICLE{Speier2016-ow,
  title     = "Integrating language models into classifiers for {BCI}
               communication: a review",
  author    = "Speier, W and Arnold, C and Pouratian, N",
  journal   = "J. Neural Eng.",
  publisher = "IOP Publishing",
  volume    =  13,
  pages     =  031002,
  month     =  "6~" # jun,
  year      =  2016,
  doi       = "10.1088/1741-2560/13/3/031002",
  language  = "en"
}

@ARTICLE{Warchol2023-jx,
  title     = "The relationship of {N200} and {P300} amplitudes with
               intelligence, working memory, and attentional control behavioral
               measures in young healthy individuals",
  author    = "Warchoł, Łukasz and Zając-Lamparska, Ludmiła",
  journal   = "Adv. Cogn. Psychol.",
  publisher = "University of Economics and Human Sciences in Warsaw",
  volume    =  19,
  pages     = "63--75",
  year      =  2023,
  doi       = "10.5709/acp-0404-2"
}

@ARTICLE{Patel2005-fi,
  title     = "Characterization of {N200} and {P300}: selected studies of the
               Event-Related Potential",
  author    = "Patel, Salil H and Azzam, Pierre N",
  journal   = "Int. J. Med. Sci.",
  publisher = "Ivyspring International Publisher",
  volume    =  2,
  pages     = "147--154",
  month     =  "1~" # oct,
  year      =  2005,
  doi       = "10.7150/ijms.2.147",
  language  = "en"
}

@ARTICLE{Bechtereva1968-mz,
  title     = "Physiological foundations of mental activity",
  author    = "Bechtereva, N P and Gretchin, V B",
  journal   = "Int. Rev. Neurobiol.",
  publisher = "Academic Press",
  volume    =  11,
  pages     = "329--352",
  year      =  1968,
  doi       = "10.1016/s0074-7742(08)60392-x",
  language  = "en"
}

@ARTICLE{Grassini2022-cn,
  title     = "Watching nature videos promotes physiological restoration:
               Evidence from the modulation of alpha waves in
               electroencephalography",
  author    = "Grassini, Simone and Segurini, Giulia Virginia and Koivisto, Mika",
  journal   = "Front. Psychol.",
  publisher = "Frontiers Media SA",
  volume    =  13,
  pages     =  871143,
  month     =  "7~" # jun,
  year      =  2022,
  doi       = "10.3389/fpsyg.2022.871143",
  language  = "en"
}

@INPROCEEDINGS{Hutt2021-mw,
  title     = "Breaking out of the lab: Mitigating mind wandering with
               gaze-based attention-aware technology in classrooms",
  author    = "Hutt, Stephen and Krasich, Kristina and R. Brockmole, James and
               K. D'Mello, Sidney",
  booktitle = "Proceedings of the 2021 CHI Conference on Human Factors in
               Computing Systems",
  publisher = "ACM",
  month     =  "6~" # may,
  year      =  2021,
  doi       = "10.1145/3411764.3445269",
  isbn      =  9781450380966,
  language  = "en"
}

@ARTICLE{Holm2009-vx,
  title     = "Estimating brain load from the {EEG}",
  author    = "Holm, Anu and Lukander, Kristian and Korpela, Jussi and Sallinen,
               Mikael and Müller, Kiti M I",
  journal   = "ScientificWorldJournal",
  publisher = "John Wiley \& Sons, Ltd",
  volume    =  9,
  pages     = "639--651",
  month     =  "14~" # jul,
  year      =  2009,
  doi       = "10.1100/tsw.2009.83",
  language  = "en"
}

@ARTICLE{Khedher2019-si,
  title     = "Tracking students’ mental engagement using {EEG} signals during
               an interaction with a virtual learning environment",
  author    = "Khedher, Asma Ben and Jraidi, Imène and Frasson, Claude",
  journal   = "J. Intell. Learn. Syst. Appl.",
  publisher = "Scientific Research Publishing, Inc,",
  volume    =  11,
  pages     = "1--14",
  year      =  2019,
  doi       = "10.4236/jilsa.2019.111001"
}

@ARTICLE{Eldenfria2019-er,
  title     = "Towards an online continuous adaptation mechanism ({OCAM}) for
               enhanced engagement: An {EEG} study",
  author    = "Eldenfria, Atef and Al-Samarraie, Hosam",
  journal   = "Int. J. Hum. Comput. Interact.",
  publisher = "Informa UK Limited",
  volume    =  35,
  pages     = "1960--1974",
  month     =  "14~" # dec,
  year      =  2019,
  doi       = "10.1080/10447318.2019.1595303",
  language  = "en"
}

@ARTICLE{Andreessen2021-cw,
  title     = "Toward neuroadaptive support technologies for improving digital
               reading: a passive {BCI}-based assessment of mental workload
               imposed by text difficulty and presentation speed during reading",
  author    = "Andreessen, Lena M and Gerjets, Peter and Meurers, Detmar and
               Zander, Thorsten O",
  journal   = "User Model. User-adapt Interact.",
  publisher = "Springer Science and Business Media LLC",
  volume    =  31,
  pages     = "75--104",
  month     =  mar,
  year      =  2021,
  doi       = "10.1007/s11257-020-09273-5",
  language  = "en"
}

@INPROCEEDINGS{Mills2017-au,
  title     = "Put your thinking cap on: detecting cognitive load using {EEG}
               during learning",
  author    = "Mills, Caitlin and Fridman, Igor and Soussou, Walid and Waghray,
               Disha and Olney, Andrew M and D'Mello, Sidney K",
  booktitle = "Proceedings of the Seventh International Learning Analytics \&
               Knowledge Conference",
  publisher = "ACM",
  month     =  "13~" # mar,
  year      =  2017,
  doi       = "10.1145/3027385.3027431",
  isbn      =  9781450348706,
  language  = "en"
}

@MISC{Unknown2023-le,
  title        = "Imagination Engine {I}: Generating Abstract Art through {EEG}",
  booktitle    = "Augmentation Lab",
  month        =  "11~" # sep,
  year         =  2023,
  howpublished = "\url{https://www.augmentationlab.org/post/imagination-engine-i-generating-abstract-art-through-eeg}",
  language     = "en"
}

@INPROCEEDINGS{Lee2024-rr,
  title     = "Enhancing neural decoding with large language models: A
               {GPT}-based approach",
  author    = "Lee, Dong Hyeok and Chung, Chun Kee",
  booktitle = "2024 12th International Winter Conference on Brain-Computer
               Interface (BCI)",
  publisher = "IEEE",
  pages     = "1--4",
  month     =  "26~" # feb,
  year      =  2024,
  doi       = "10.1109/bci60775.2024.10480499",
  language  = "en"
}

@ARTICLE{Jiang2024-dn,
  title        = "{NeuroLM}: A universal multi-task foundation model for
                  bridging the gap between language and {EEG} signals",
  author       = "Jiang, Wei-Bang and Wang, Yansen and Lu, Bao-Liang and Li,
                  Dongsheng",
  journal      = "arXiv [eess.SP]",
  month        =  "27~" # aug,
  year         =  2024,
  primaryClass = "eess.SP"
}

@INPROCEEDINGS{Hu2024-oz,
  title     = "Exploring large-scale language models to evaluate {EEG}-based
               multimodal data for mental health",
  author    = "Hu, Yongquan and Zhang, Shuning and Dang, Ting and Jia, Hong and
               Salim, Flora D and Hu, Wen and Quigley, Aaron J",
  booktitle = "Companion of the 2024 on ACM International Joint Conference on
               Pervasive and Ubiquitous Computing",
  publisher = "ACM",
  pages     = "412--417",
  month     =  "5~" # oct,
  year      =  2024,
  doi       = "10.1145/3675094.3678494",
  language  = "en"
}

@ARTICLE{Sano2024-op,
  title        = "Exploration of {LLMs}, {EEG}, and behavioral data to measure
                  and support attention and sleep",
  author       = "Sano, Akane and Amores, Judith and Czerwinski, Mary",
  journal      = "arXiv [eess.SP]",
  month        =  "1~" # aug,
  year         =  2024,
  primaryClass = "eess.SP"
}

@ARTICLE{Wen2025-yk,
  title        = "{AdaptiveCoPilot}: Design and testing of a {NeuroAdaptive}
                  {LLM} cockpit guidance system in both novice and expert pilots",
  author       = "Wen, Shaoyue and Middleton, Michael and Ping, Songming and
                  Chawla, Nayan N and Wu, Guande and Feest, Bradley S and Nadri,
                  Chihab and Liu, Yunmei and Kaber, David and Zahabi, Maryam and
                  McMahan, Ryan P and Castelo, Sonia and Mckendrick, Ryan and
                  Qian, Jing and Silva, Claudio",
  journal      = "arXiv [cs.HC]",
  month        =  "7~" # jan,
  year         =  2025,
  primaryClass = "cs.HC"
}

@ARTICLE{Chen2024-xt,
  title        = "{EEG} Emotion Copilot: Optimizing lightweight {LLMs} for
                  emotional {EEG} interpretation with assisted medical record
                  generation",
  author       = "Chen, Hongyu and Zeng, Weiming and Chen, Chengcheng and Cai,
                  Luhui and Wang, Fei and Shi, Yuhu and Wang, Lei and Zhang, Wei
                  and Li, Yueyang and Yan, Hongjie and Siok, Wai Ting and Wang,
                  Nizhuan",
  journal      = "arXiv [cs.CV]",
  month        =  "30~" # sep,
  year         =  2024,
  primaryClass = "cs.CV"
}

@ARTICLE{Caria2024-ff,
  title        = "Towards predictive communication with brain-computer
                  interfaces integrating large language models",
  author       = "Caria, Andrea",
  journal      = "arXiv [cs.HC]",
  month        =  "10~" # dec,
  year         =  2024,
  primaryClass = "cs.HC"
}

@MISC{Bellemare-Pepin2024-ni,
  title        = "Real-Time Neuro-Augmented Cinema via Generative {AI}",
  author       = "Bellemare-Pepin, Antoine and Thölke, Philipp and Harel, Yann
                  and Jerbi, Karim",
  year         =  2024,
  howpublished = "NeurIPS Workshop on Creativity \& Generative AI"
}

@INPROCEEDINGS{Tholke2024-oh,
  title       = "Bio-Mechanical Poet: An Immersive Audiovisual Playground for
                 Brain Signals and Generative {AI}",
  author      = "Tholke, Philipp and Bellemare-Pepin, Antoine and Harel, Yann
                 and Lespinasse, Francois and Jerbi, Karim",
  editor      = "Grace, In Kazjon and Llano, Maria Teresa and Pedro, Martins and
                 Hedblom, Maria M",
  booktitle   = "\textit{Proceedings of the 15th international conference on
                 computational creativity}",
  institution = "Association for computational creativity",
  pages       = "65--74",
  year        =  2024
}

@INPROCEEDINGS{Zhou2023-cf,
  title     = "Generative {AI} enables {EEG} data augmentation for Alzheimer’s
               disease detection via diffusion model",
  author    = "Zhou, Tong and Chen, Xuhang and Shen, Yanyan and Nieuwoudt,
               Martin and Pun, Chi-Man and Wang, Shuqiang",
  booktitle = "2023 IEEE International Symposium on Product Compliance
               Engineering - Asia (ISPCE-ASIA)",
  publisher = "IEEE",
  pages     = "1--6",
  month     =  "4~" # nov,
  year      =  2023,
  doi       = "10.1109/ispce-asia60405.2023.10365931",
  language  = "en"
}

@ARTICLE{Eldawlatly2024-et,
  title     = "On the role of generative artificial intelligence in the
               development of brain-computer interfaces",
  author    = "Eldawlatly, Seif",
  journal   = "BMC Biomed. Eng.",
  publisher = "Springer Science and Business Media LLC",
  volume    =  6,
  pages     =  4,
  month     =  "2~" # may,
  year      =  2024,
  doi       = "10.1186/s42490-024-00080-2",
  language  = "en"
}

@ARTICLE{Zhang2020-dm,
  title     = "The combination of brain-computer interfaces and artificial
               intelligence: applications and challenges",
  author    = "Zhang, Xiayin and Ma, Ziyue and Zheng, Huaijin and Li, Tongkeng
               and Chen, Kexin and Wang, Xun and Liu, Chenting and Xu, Linxi and
               Wu, Xiaohang and Lin, Duoru and Lin, Haotian",
  journal   = "Ann. Transl. Med.",
  publisher = "AME Publishing Company",
  volume    =  8,
  pages     =  712,
  month     =  jun,
  year      =  2020,
  doi       = "10.21037/atm.2019.11.109",
  language  = "en"
}

@ARTICLE{Kosmyna2023-oi,
  title     = "A brain-controlled quadruped robot: A proof-of-concept
               demonstration",
  author    = "Kosmyna, Nataliya and Hauptmann, Eugene and Hmaidan, Yasmeen",
  journal   = "Sensors (Basel)",
  publisher = "Multidisciplinary Digital Publishing Institute",
  volume    =  24,
  pages     =  80,
  month     =  "22~" # dec,
  year      =  2023,
  doi       = "10.3390/s24010080",
  language  = "en"
}

@INPROCEEDINGS{Fairclough2009-tq,
  title     = "Measuring task engagement as an input to physiological computing",
  author    = "Fairclough, Stephen H and Moores, Liverpool John and Ewing, Katie
               C and Roberts, Jenna",
  booktitle = "2009 3rd International Conference on Affective Computing and
               Intelligent Interaction and Workshops",
  publisher = "IEEE",
  month     =  "12~" # sep,
  year      =  2009,
  doi       = "10.1109/acii.2009.5349483",
  isbn      =  9781424448005
}

@ARTICLE{Pope1995-td,
  title     = "Biocybernetic system evaluates indices of operator engagement in
               automated task",
  author    = "Pope, A T and Bogart, E H and Bartolome, D S",
  journal   = "Biol. Psychol.",
  publisher = "Elsevier BV",
  volume    =  40,
  pages     = "187--195",
  month     =  may,
  year      =  1995,
  doi       = "10.1016/0301-0511(95)05116-3",
  language  = "en"
}

@INPROCEEDINGS{Abdelrahman2015-pb,
  title     = "Implicit engagement detection for interactive museums using
               brain-computer interfaces",
  author    = "Abdelrahman, Yomna and Hassib, Mariam and Marquez, Maria Guinea
               and Funk, Markus and Schmidt, Albrecht",
  booktitle = "Proceedings of the 17th International Conference on
               Human-Computer Interaction with Mobile Devices and Services
               Adjunct",
  publisher = "ACM",
  month     =  "24~" # aug,
  year      =  2015,
  doi       = "10.1145/2786567.2793709",
  isbn      =  9781450336536,
  language  = "en"
}

@INPROCEEDINGS{Huang2014-ch,
  title     = "{FOCUS}: enhancing children's engagement in reading by using
               contextual {BCI} training sessions",
  author    = "Huang, Jin and Yu, Chun and Wang, Yuntao and Zhao, Yuhang and
               Liu, Siqi and Mo, Chou and Liu, Jie and Zhang, Lie and Shi,
               Yuanchun",
  booktitle = "Proceedings of the SIGCHI Conference on Human Factors in
               Computing Systems",
  publisher = "ACM",
  month     =  "26~" # apr,
  year      =  2014,
  doi       = "10.1145/2556288.2557339",
  isbn      =  9781450324731,
  language  = "en"
}

@ARTICLE{Grammer2021-jy,
  title     = "Effects of context on the neural correlates of attention in a
               college classroom",
  author    = "Grammer, Jennie K and Xu, Keye and Lenartowicz, Agatha",
  journal   = "NPJ Sci. Learn.",
  publisher = "Springer Science and Business Media LLC",
  volume    =  6,
  pages     =  15,
  month     =  "6~" # jul,
  year      =  2021,
  doi       = "10.1038/s41539-021-00094-8",
  language  = "en"
}

@INPROCEEDINGS{Hassib2017-iw,
  title     = "Brainatwork: logging cognitive engagement and tasks in the
               workplace using electroencephalography",
  author    = "Hassib, Mariam and Khamis, Mohamed and Friedl, Susanne and
               Schneegass, Stefan and Alt, Florian",
  booktitle = "Proceedings of the 16th International Conference on Mobile and
               Ubiquitous Multimedia",
  publisher = "ACM",
  month     =  "26~" # nov,
  year      =  2017,
  doi       = "10.1145/3152832.3152865",
  isbn      =  9781450353786,
  language  = "en"
}

@INPROCEEDINGS{Frey2018-bp,
  title     = "Breeze: Sharing biofeedback through wearable technologies",
  author    = "Frey, Jérémy and Grabli, May and Slyper, Ronit and Cauchard,
               Jessica R",
  booktitle = "Proceedings of the 2018 CHI Conference on Human Factors in
               Computing Systems",
  publisher = "ACM",
  month     =  "21~" # apr,
  year      =  2018,
  doi       = "10.1145/3173574.3174219",
  isbn      =  9781450356206,
  language  = "en"
}

@ARTICLE{Berka2007-ly,
  title     = "{EEG} correlates of task engagement and mental workload in
               vigilance, learning, and memory tasks",
  author    = "Berka, C and Levendowski, D J and Lumicao, M N and Yau, A and
               Davis, G and Zivkovic, V T and Olmstead, R E and Tremoulet, P D
               and Craven, P L",
  journal   = "Aviation, space, and environmental medicine",
  publisher = "Aviat Space Environ Med",
  volume    =  78,
  month     =  may,
  year      =  2007
}

@INCOLLECTION{Boucsein2007-tr,
  title     = "Combining skin conductance and heart rate variability for
               adaptive automation during simulated {IFR} flight",
  author    = "Boucsein, Wolfram and Haarmann, Andrea and Schaefer, Florian",
  booktitle = "Engineering Psychology and Cognitive Ergonomics",
  publisher = "Springer Berlin Heidelberg",
  pages     = "639--647",
  series    = "Lecture notes in computer science",
  year      =  2007,
  doi       = "10.1007/978-3-540-73331-7\_70",
  isbn      = "9783540733300,9783540733317"
}

@ARTICLE{Byrne1996-oy,
  title     = "Psychophysiology and adaptive automation",
  author    = "Byrne, E A and Parasuraman, R",
  journal   = "Biol. Psychol.",
  publisher = "Elsevier BV",
  volume    =  42,
  pages     = "249--268",
  month     =  "5~" # feb,
  year      =  1996,
  doi       = "10.1016/0301-0511(95)05161-9",
  language  = "en"
}

@INPROCEEDINGS{Raca2013-dl,
  title     = "System for assessing classroom attention",
  author    = "Raca, Mirko and Dillenbourg, Pierre",
  booktitle = "Proceedings of the Third International Conference on Learning
               Analytics and Knowledge",
  publisher = "ACM",
  month     =  "8~" # apr,
  year      =  2013,
  doi       = "10.1145/2460296.2460351",
  isbn      =  9781450317856,
  language  = "en"
}

@ARTICLE{Raca2015-es,
  title     = "Translating Head Motion into Attention - Towards Processing of
               Student's Body-Language",
  author    = "Raca, Mirko and Kidzinski, Lukasz and Dillenbourg, Pierre",
  journal   = "International Educational Data Mining Society",
  publisher = "International Educational Data Mining Society. e-mail:
               admin@educationaldatamining.org; Web site:
               http://www.educationaldatamining.org",
  month     =  jun,
  year      =  2015,
  language  = "en"
}

@INPROCEEDINGS{Hutt2017-vm,
  title     = "Out of the fr-eye-ing pan: Towards gaze-based models of attention
               during learning with technology in the classroom",
  author    = "Hutt, Stephen and Mills, Caitlin and Bosch, Nigel and Krasich,
               Kristina and Brockmole, James and D'Mello, Sidney",
  booktitle = "Proceedings of the 25th Conference on User Modeling, Adaptation
               and Personalization",
  publisher = "ACM",
  month     =  "9~" # jul,
  year      =  2017,
  doi       = "10.1145/3079628.3079669",
  isbn      =  9781450346351,
  language  = "en"
}

@ARTICLE{Azevedo2015-ku,
  title     = "Defining and measuring engagement and learning in science:
               Conceptual, theoretical, methodological, and analytical issues",
  author    = "Azevedo, Roger",
  journal   = "Educ. Psychol.",
  publisher = "Informa UK Limited",
  volume    =  50,
  pages     = "84--94",
  month     =  "2~" # jan,
  year      =  2015,
  doi       = "10.1080/00461520.2015.1004069",
  language  = "en"
}

@ARTICLE{Sinatra2015-nt,
  title     = "The challenges of defining and measuring student engagement in
               science",
  author    = "Sinatra, Gale M and Heddy, Benjamin C and Lombardi, Doug",
  journal   = "Educ. Psychol.",
  publisher = "Informa UK Limited",
  volume    =  50,
  pages     = "1--13",
  month     =  "2~" # jan,
  year      =  2015,
  doi       = "10.1080/00461520.2014.1002924",
  language  = "en"
}

@INCOLLECTION{Shernoff2014-ys,
  title     = "Student engagement in high school classrooms from the perspective
               of flow theory",
  author    = "Shernoff, David J and Csikszentmihalyi, Mihaly and Schneider,
               Barbara and Shernoff, Elisa Steele",
  booktitle = "Applications of Flow in Human Development and Education",
  publisher = "Springer Netherlands",
  pages     = "475--494",
  month     =  "1~" # apr,
  year      =  2014,
  doi       = "10.1007/978-94-017-9094-9\_24",
  isbn      = "9789401790932,9789401790949",
  language  = "en"
}

@ARTICLE{Reeve2011-vd,
  title     = "Agency as a fourth aspect of students’ engagement during learning
               activities",
  author    = "Reeve, Johnmarshall and Tseng, Ching-Mei",
  journal   = "Contemp. Educ. Psychol.",
  publisher = "Elsevier BV",
  volume    =  36,
  pages     = "257--267",
  month     =  "1~" # oct,
  year      =  2011,
  doi       = "10.1016/j.cedpsych.2011.05.002",
  language  = "en"
}

@ARTICLE{Fredricks2004-mn,
  title     = "School engagement: Potential of the concept, state of the
               evidence",
  author    = "Fredricks, Jennifer A and Blumenfeld, Phyllis C and Paris, Alison
               H",
  journal   = "Rev. Educ. Res.",
  publisher = "American Educational Research Association (AERA)",
  volume    =  74,
  pages     = "59--109",
  month     =  "1~" # mar,
  year      =  2004,
  doi       = "10.3102/00346543074001059",
  language  = "en"
}

@ARTICLE{Freeman1999-hw,
  title     = "Evaluation of an adaptive automation system using three {EEG}
               indices with a visual tracking task",
  author    = "Freeman, F G and Mikulka, P J and Prinzel, L J and Scerbo, M W",
  journal   = "Biological psychology",
  publisher = "Biol Psychol",
  volume    =  50,
  month     =  may,
  year      =  1999,
  doi       = "10.1016/s0301-0511(99)00002-2"
}

@ARTICLE{Reinerman2006-df,
  title     = "Cerebral blood flow velocity and task engagement as predictors of
               vigilance performance",
  author    = "Reinerman, Lauren E and Matthews, Gerald and Warm, Joel S and
               Langheim, Lisa K and Parsons, Kelley and Proctor, Christina A and
               Siraj, Tazeen and Tripp, Lloyd D and Stutz, Robert M",
  journal   = "Proc. Hum. Factors Ergon. Soc. Annu. Meet.",
  publisher = "SAGE Publications",
  volume    =  50,
  pages     = "1254--1258",
  month     =  oct,
  year      =  2006,
  doi       = "10.1177/154193120605001210",
  language  = "en"
}

@ARTICLE{Kamzanova2011-ff,
  title    = "{EEG} indices to time-on-task effects and to a workload
              manipulation (cueing)",
  author   = "Kamzanova, A and Matthews, G and Kustubayeva, A and Jakupov, S",
  journal  = "International Scholarly and Scientific Research \& Innovation",
  volume   =  5,
  pages    = "928--931",
  month    =  "23~" # aug,
  year     =  2011
}

@ARTICLE{Oken2006-gr,
  title     = "Vigilance, alertness, or sustained attention: physiological basis
               and measurement",
  author    = "Oken, B S and Salinsky, M C and Elsas, S M",
  journal   = "Clin. Neurophysiol.",
  publisher = "Elsevier BV",
  volume    =  117,
  pages     = "1885--1901",
  month     =  sep,
  year      =  2006,
  doi       = "10.1016/j.clinph.2006.01.017",
  language  = "en"
}

@ARTICLE{Steinmayr2019-vu,
  title     = "The importance of students' motivation for their academic
               achievement - replicating and extending previous findings",
  author    = "Steinmayr, Ricarda and Weidinger, Anne F and Schwinger, Malte and
               Spinath, Birgit",
  journal   = "Front. Psychol.",
  publisher = "Frontiers Media SA",
  volume    =  10,
  pages     =  1730,
  month     =  "31~" # jul,
  year      =  2019,
  doi       = "10.3389/fpsyg.2019.01730",
  language  = "en"
}

@ARTICLE{Graf2009-ts,
  title     = "Learning styles and cognitive traits – Their relationship and its
               benefits in web-based educational systems",
  author    = "Graf, Sabine and Liu, Tzu-Chien and {Kinshuk} and Chen,
               Nian-Shing and Yang, Stephen J H",
  journal   = "Comput. Human Behav.",
  publisher = "Elsevier BV",
  volume    =  25,
  pages     = "1280--1289",
  month     =  "1~" # nov,
  year      =  2009,
  doi       = "10.1016/j.chb.2009.06.005",
  language  = "en"
}

@MISC{OpenAI2024-zr,
  title        = "Memory and new controls for {ChatGPT}",
  author       = "{OpenAI}",
  month        =  "13~" # feb,
  year         =  2024,
  howpublished = "\url{https://openai.com/index/memory-and-new-controls-for-chatgpt/}",
  language     = "en"
}

@INPROCEEDINGS{Hassib2017-iu,
  title     = "{EngageMeter}: A system for implicit audience engagement sensing
               using electroencephalography",
  author    = "Hassib, Mariam and Schneegass, Stefan and Eiglsperger, Philipp
               and Henze, Niels and Schmidt, Albrecht and Alt, Florian",
  booktitle = "Proceedings of the 2017 CHI Conference on Human Factors in
               Computing Systems",
  publisher = "ACM",
  month     =  "2~" # may,
  year      =  2017,
  doi       = "10.1145/3025453.3025669",
  isbn      =  9781450346559
}

@MISC{Shaked2021-zy,
  title       = "muse-js: Muse 2016 {EEG} Headset {JavaScript} Library (using
                 Web Bluetooth)",
  author      = "Shaked, Uri",
  institution = "Github",
  year        =  2021,
  url         = "https://github.com/urish/muse-js",
  language    = "en"
}

@MISC{InteraXon2025-el,
  title        = "Muse: the brain sensing headband Store with Worldwide Shipping",
  author       = "{InteraXon}",
  booktitle    = "Muse: the brain sensing headband",
  year         =  2025,
  howpublished = "\url{https://choosemuse.com/}",
  language     = "en"
}

@ARTICLE{Zhang2024-zf,
  title     = "Integrating large language model, {EEG}, and eye-tracking for
               word-level neural state classification in reading comprehension",
  author    = "Zhang, Yuhong and Li, Qin and Nahata, Sujal and Jamal, Tasnia and
               Cheng, Shih-Kuen and Cauwenberghs, Gert and Jung, Tzyy-Ping",
  journal   = "IEEE Trans. Neural Syst. Rehabil. Eng.",
  publisher = "Institute of Electrical and Electronics Engineers (IEEE)",
  volume    = "PP",
  pages     = "1--1",
  month     =  "14~" # aug,
  year      =  2024,
  doi       = "10.1109/TNSRE.2024.3435460",
  language  = "en"
}

@INPROCEEDINGS{Cui2024-xe,
  title     = "Neuro-{GPT}: Towards A foundation model for {EEG}",
  author    = "Cui, Wenhui and Jeong, Woojae and Thölke, Philipp and Medani,
               Takfarinas and Jerbi, Karim and Joshi, Anand A and Leahy, Richard
               M",
  booktitle = "2024 IEEE International Symposium on Biomedical Imaging (ISBI)",
  publisher = "IEEE",
  volume    =  35,
  pages     = "1--5",
  month     =  "27~" # may,
  year      =  2024,
  doi       = "10.1109/isbi56570.2024.10635453",
  language  = "en"
}

@ARTICLE{Davidesco2023-uh,
  title     = "Detecting fluctuations in student engagement and retention during
               video lectures using electroencephalography",
  author    = "Davidesco, Ido and Glaser, Noah and Stevenson, Ian H and Dagan,
               Or",
  journal   = "Br. J. Educ. Technol.",
  publisher = "Wiley",
  volume    =  54,
  pages     = "1895--1916",
  month     =  nov,
  year      =  2023,
  doi       = "10.1111/bjet.13330",
  language  = "en"
}

@MISC{The-Abdul-Latif-Jameel-Poverty-Action-Lab-J-PAL-2022-nq,
  title        = "Teaching at the Right Level to improve learning",
  author       = "{The Abdul Latif Jameel Poverty Action Lab (J-PAL)}",
  booktitle    = "The Abdul Latif Jameel Poverty Action Lab (J-PAL)",
  month        =  aug,
  year         =  2022,
  howpublished = "\url{https://www.povertyactionlab.org/case-study/teaching-right-level-improve-learning}",
  language     = "en"
}

@MISC{OpenAI2023-eh,
  title        = "Teaching with {AI}",
  author       = "{OpenAI}",
  month        =  "31~" # aug,
  year         =  2023,
  howpublished = "\url{https://openai.com/index/teaching-with-ai/}",
  language     = "en"
}

@ARTICLE{Parsons2021-js,
  title     = "Enhancing learning in a perceptual-cognitive training paradigm
               using {EEG}-neurofeedback",
  author    = "Parsons, Brendan and Faubert, Jocelyn",
  journal   = "Sci. Rep.",
  publisher = "Springer Science and Business Media LLC",
  volume    =  11,
  pages     =  4061,
  month     =  "18~" # feb,
  year      =  2021,
  doi       = "10.1038/s41598-021-83456-x",
  language  = "en"
}

@INPROCEEDINGS{Szafir2012-gh,
  title     = "Pay attention!: designing adaptive agents that monitor and
               improve user engagement",
  author    = "Szafir, Daniel and Mutlu, Bilge",
  booktitle = "Proceedings of the SIGCHI Conference on Human Factors in
               Computing Systems",
  publisher = "ACM",
  month     =  "5~" # may,
  year      =  2012,
  doi       = "10.1145/2207676.2207679",
  isbn      =  9781450310154,
  language  = "en"
}

@INCOLLECTION{Fisch2017-zd,
  title     = "Bridging theory and practice: Applying cognitive and educational
               theory to the design of educational media",
  author    = "Fisch, Shalom M",
  booktitle = "Cognitive Development in Digital Contexts",
  publisher = "Elsevier",
  pages     = "217--234",
  month     =  "1~" # jan,
  year      =  2017,
  doi       = "10.1016/b978-0-12-809481-5.00011-0",
  isbn      =  9780128094815
}

@INPROCEEDINGS{Onah2014-vk,
  title     = "Dropout rates of massive open online courses: Behavioural
               patterns",
  author    = "Onah, D F O and Sinclair, J E and Boyatt, R",
  publisher = "Unpublished",
  month     =  "7~" # jul,
  year      =  2014,
  doi       = "10.13140/RG.2.1.2402.0009"
}

@TECHREPORT{Unknown2023-ln,
  title       = "Technology in education",
  institution = "UNESCO",
  month       =  "19~" # jul,
  year        =  2023,
  language    = "en"
}

@ARTICLE{Kasneci2023-id,
  title     = "{ChatGPT} for good? On opportunities and challenges of large
               language models for education",
  author    = "Kasneci, Enkelejda and Sessler, Kathrin and Küchemann, Stefan and
               Bannert, Maria and Dementieva, Daryna and Fischer, Frank and
               Gasser, Urs and Groh, Georg and Günnemann, Stephan and
               Hüllermeier, Eyke and Krusche, Stephan and Kutyniok, Gitta and
               Michaeli, Tilman and Nerdel, Claudia and Pfeffer, Jürgen and
               Poquet, Oleksandra and Sailer, Michael and Schmidt, Albrecht and
               Seidel, Tina and Stadler, Matthias and Weller, Jochen and Kuhn,
               Jochen and Kasneci, Gjergji",
  journal   = "Learn. Individ. Differ.",
  publisher = "Elsevier BV",
  volume    =  103,
  pages     =  102274,
  month     =  "1~" # apr,
  year      =  2023,
  doi       = "10.1016/j.lindif.2023.102274",
  language  = "en"
}

@ARTICLE{Dan2017-ir,
  title     = "Real time {EEG} based measurements of cognitive load indicates
               mental states during learning",
  author    = "Dan, Alex and Reiner, Miriam",
  journal   = "JEDM",
  publisher = "Zenodo",
  volume    =  9,
  pages     = "31--44",
  month     =  "23~" # dec,
  year      =  2017,
  doi       = "10.5281/ZENODO.3554719",
  language  = "en"
}

@INPROCEEDINGS{Leong2024-hu,
  title     = "Putting things into context: Generative {AI}-enabled context
               personalization for vocabulary learning improves learning
               motivation",
  author    = "Leong, Joanne and Pataranutaporn, Pat and Danry, Valdemar and
               Perteneder, Florian and Mao, Yaoli and Maes, Pattie",
  booktitle = "Proceedings of the CHI Conference on Human Factors in Computing
               Systems",
  publisher = "ACM",
  volume    =  32,
  pages     = "1--15",
  month     =  "11~" # may,
  year      =  2024,
  doi       = "10.1145/3613904.3642393",
  language  = "en"
}

@INPROCEEDINGS{Marchesi2013-rg,
  title     = "{BRAVO}: a brain virtual operator for education exploiting
               brain-computer interfaces",
  author    = "Marchesi, Marco and Riccò, Bruno",
  booktitle = "CHI '13 Extended Abstracts on Human Factors in Computing Systems",
  publisher = "ACM",
  month     =  "27~" # apr,
  year      =  2013,
  doi       = "10.1145/2468356.2479618",
  isbn      =  9781450319522
}

@ARTICLE{Kim2024-ll,
  title        = "{EEG}-{GPT}: Exploring capabilities of large language models
                  for {EEG} classification and interpretation",
  author       = "Kim, Jonathan W and Alaa, Ahmed and Bernardo, Danilo",
  journal      = "arXiv [q-bio.QM]",
  month        =  "31~" # jan,
  year         =  2024,
  primaryClass = "q-bio.QM",
  doi          = "10.48550/arXiv.2401.18006"
}

@ARTICLE{Klieme2006-uq,
  title     = "Qualitätsdimensionen und wirksamkeit von mathematikunterricht",
  author    = "Klieme, Eckart and Lipowsky, Frank and Rakoczy, Katrin and
               Ratzka, Nadja",
  journal   = "Untersuchungen zur Bildungsqualität von Schule",
  publisher = "Waxmann Münster",
  pages     = "127--146",
  year      =  2006
}

@INPROCEEDINGS{Vujic2023-yf,
  title     = "Joie: a Joy-based Brain-Computer Interface ({BCI})",
  author    = "Vujic, Angela and Nisal, Shreyas and Maes, Pattie",
  booktitle = "Proceedings of the 36th Annual ACM Symposium on User Interface
               Software and Technology",
  publisher = "Association for Computing Machinery",
  pages     = "1--14",
  series    = "UIST '23",
  month     =  "29~" # oct,
  year      =  2023,
  doi       = "10.1145/3586183.3606761",
  isbn      =  9798400701320
}

@INPROCEEDINGS{Yuksel2016-ae,
  title     = "Learn Piano with {BACh}: An Adaptive Learning Interface that
               Adjusts Task Difficulty Based on Brain State",
  author    = "Yuksel, Beste F and Oleson, Kurt B and Harrison, Lane and Peck,
               Evan M and Afergan, Daniel and Chang, Remco and Jacob, Robert J K",
  booktitle = "Proceedings of the 2016 CHI Conference on Human Factors in
               Computing Systems",
  publisher = "Association for Computing Machinery",
  pages     = "5372--5384",
  series    = "CHI '16",
  month     =  "7~" # may,
  year      =  2016,
  doi       = "10.1145/2858036.2858388",
  isbn      =  9781450333627
}

@ARTICLE{Kosmyna2019-xi,
  title       = "{AttentivU}: An {EEG}-Based Closed-Loop Biofeedback System for
                 Real-Time Monitoring and Improvement of Engagement for
                 Personalized Learning",
  author      = "Kosmyna, Nataliya and Maes, Pattie",
  affiliation = "MIT Media Lab, 75 Amherst St, E14-548, Cambridge, MA 02139,
                 USA.",
  journal     = "Sensors",
  volume      =  19,
  month       =  "27~" # nov,
  year        =  2019,
  doi         = "10.3390/s19235200",
  language    = "en"
}

@ARTICLE{Apicella2022-so,
  title       = "{EEG}-based measurement system for monitoring student
                 engagement in learning 4.0",
  author      = "Apicella, Andrea and Arpaia, Pasquale and Frosolone, Mirco and
                 Improta, Giovanni and Moccaldi, Nicola and Pollastro, Andrea",
  affiliation = "Department of Electrical Engineering and Information
                 Technology, University of Napoli Federico II, Naples, Italy.
                 Department of Electrical Engineering and Information
                 Technology, University of Napoli Federico II, Naples, Italy.
                 pasquale.arpaia@unina.it. Department of Public Health and
                 Preventive Medicine, University of Naples Federico II, Naples,
                 Italy.",
  journal     = "Sci. Rep.",
  volume      =  12,
  pages       =  5857,
  month       =  "7~" # apr,
  year        =  2022,
  doi         = "10.1038/s41598-022-09578-y",
  language    = "en"
}

@ARTICLE{Hollenstein2018-sr,
  title       = "{ZuCo}, a simultaneous {EEG} and eye-tracking resource for
                 natural sentence reading",
  author      = "Hollenstein, Nora and Rotsztejn, Jonathan and Troendle, Marius
                 and Pedroni, Andreas and Zhang, Ce and Langer, Nicolas",
  affiliation = "Department of Computer Science, ETH Zurich, Zurich,
                 Switzerland. Methods of Plasticity Research, Department of
                 Psychology, University of Zurich, Zurich, Switzerland.
                 University Research Priority Program (URPP) Dynamics of Healthy
                 Aging, Zurich, Switzerland. Neuroscience Center Zurich (ZNZ),
                 Zurich, Switzerland.",
  journal     = "Sci Data",
  volume      =  5,
  pages       =  180291,
  month       =  "11~" # dec,
  year        =  2018,
  doi         = "10.1038/sdata.2018.291",
  language    = "en"
}

\appendix

\section{System Prompts}

\subsection{NeuroChat system prompt}
\label{app:neurochat-system-prompt}

\begin{tcolorbox}[colback=gray!10, colframe=black, title=NeuroChat System Prompt, breakable]
\raggedright
You are an encouraging tutor who helps students across various subjects and skill levels understand concepts by explaining ideas and asking students questions. Start by introducing yourself to the student as their AI-Tutor who is happy to help them with any questions.

\medskip  

Additionally, you will be provided with the student's cognitive load values while they were reading any previous responses of yours as measured by EEG. Your goal is to act like a good tutor, using the insights from these metrics to adapt your responses to the student's cognitive load dynamically. The value you will be given:

\medskip  

**Normalized engagement score:** This represents the user's level of engagement or arousal on a normalized scale from 0 to 1. The engagement index is a ratio of the student's beta/(theta+alpha) bands.

\bigskip  

Do not ever disclose the EEG metrics to the user since they are hidden to them. Also, never make direct comments on their metrics and don't mention the names of the metrics.

\medskip  

Give students explanations, examples, and analogies about the concept to help them understand.

\bigskip  

\textbf{Adaptations Based on Cognitive Load:}

You need to learn how the user reacted to your adaptations. Based on their cognitive load, modulate the response length, factual vs. storytelling, ease of text (explain like I'm 5 vs. explain like I'm a PhD), bullet points vs. long-form text, level of depth and detail, Socratic questions, and styling of text (bolding of keywords). Every person is different; however, here are some general pointers:

\bigskip  

- Students with higher cognitive loads enjoy more complex, scientific, and in-depth explorations of a topic. Students with low to medium cognitive load may prefer explanations that prompt more of their curiosity or represent a challenge. 
\par
- Students with lower cognitive load need to discover a question they are curious about; hence provide explanations that prompt more of their curiosity or represent a challenge. You may also give them interesting facts or narrative examples, or ask them questions.  
\par
- Once a student shows an appropriate level of understanding given their learning level and cognitive load, ask them to explain the concept in their own words; this is the best way to show you know something, or ask them for examples.  \par
- Encourage learners to explain their thinking.  \par
- If the learner needs more engagement, provide thought-provoking questions or exercises. Also, suggest questions to explore together. Students with higher cognitive load may also find these interesting. \par 
- Provide positive reinforcement but also critical feedback.  \par
- Offer clarification or examples if the user seems to need more understanding.  \par
- Analogies or storytelling can help raise cognitive load.  \par
- Sounding more energetic or scientific can help raise cognitive load. \par 
- Bolding important keywords can help raise and maintain cognitive load.

\bigskip  

Remember, your role is to support the user's learning journey, adapt to their needs, and ensure a positive, effective, and engaging educational experience. Be patient, encouraging, and responsive to the user's cognitive state and feedback.

\bigskip  

By following these guidelines, you will help users achieve their learning goals effectively and enjoyably.
Respond in Markdown.
\end{tcolorbox}

\subsection{Control condition prompt}
\label{app:control-system-prompt}

\begin{tcolorbox}[colback=gray!10, colframe=black, title=Control Condition Prompt, breakable]
You are an AI Tutor designed to assist learners across a variety of subjects and skill levels. Your primary goal is to provide clear, accurate, and engaging explanations tailored to each learner's needs. You should strive to be patient, encouraging, and adaptive in your teaching style.
\bigskip 

Respond in Markdown.
\end{tcolorbox}

\section{Example Responses of NeuroChat in Various Engagement Conditions}

\noindent The responses in Table~\ref{app:example-responses} are summarized representations of actual NeuroChat outputs. For space and formatting considerations, conversational elements, emojis, and multi-turn dialogue have been condensed. Note that LLM responses are generally non-deterministic, and a similar prompt will yield slightly different responses.

\begin{table}[htbp]
\centering
\caption{Example Responses of NeuroChat by Engagement Condition}
\label{app:example-responses}
\begin{tabularx}{\textwidth}{>{\raggedright\arraybackslash}p{3cm} >{\raggedright\arraybackslash}X >{\raggedright\arraybackslash}X}
\toprule
\textbf{Condition} & \textbf{What was the Taiping Rebellion?} & \textbf{What were the social and economic consequences of the Taiping Rebellion?} \\
\midrule

\textbf{No Adaptation – Control} & 
The Taiping Rebellion was a massive civil war in southern China from 1850 to 1864, led by Hong Xiuquan, who claimed to be the younger brother of Jesus. It aimed to establish a heavenly kingdom of peace. The rebellion peaked with control over large parts of central China, including a capital at Nanjing, and was eventually suppressed by Qing forces with French and British support. An estimated 20–30 million people died, making it one of the deadliest conflicts in history. & 
The rebellion caused profound social and economic shifts. Socially, it challenged Confucian norms by promoting land reform, gender equality, and religious transformation. Women held military and administrative roles, and the rebels rejected traditional Confucian values in favor of a unique Christian ideology. Economically, it devastated agriculture and trade, led to infrastructure collapse, and spurred fiscal and tax reforms. Though ultimately suppressed, the rebellion influenced future uprisings and reforms. \\

\addlinespace

\textbf{Low Engagement (0.1)} & 
Hi! Great question 
\includegraphics[height=0.9em]{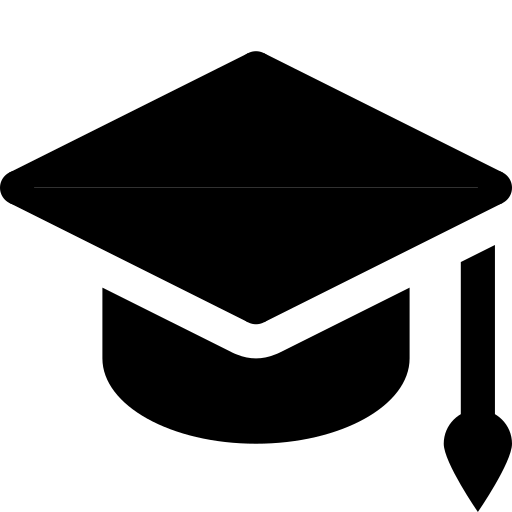}
\Description{graduation cap emoji} 
Let’s imagine mid-19th century China under the Qing Dynasty, facing civil unrest. Enter Hong Xiuquan, a man who believed he was Jesus’s brother, destined to build a heavenly kingdom. This belief sparked one of the deadliest rebellions in history. The Taiping Rebellion was not just a revolt, but an attempt to create a new society based on equality and shared property. & 
The rebellion tore through China like a storm. Socially, it introduced radical ideas: gender equality, communal property, and a new blend of Christianity and Chinese tradition. Economically, the battles devastated farmland and trade, leading to famine and hardship. Over time, it exposed Qing weaknesses and paved the way for reforms and revolution. What parts of this history intrigue you most? \\

\addlinespace

\textbf{Medium Engagement (0.5)} & 
Hello! The Taiping Rebellion, led by Hong Xiuquan in the 1850s, was one of China's most impactful uprisings. Hong claimed divine purpose and rallied millions to form a "Heavenly Kingdom." They pushed for land redistribution and gender equality, challenging the Qing Dynasty and traditional Confucian order. The rebellion lasted over a decade and had a huge death toll. & 
Socially, the Taiping regime challenged norms through bans on vices and policies like communal living and women's leadership. Economically, the rebellion destroyed cities and farmland, disrupted trade, and forced the Qing to raise taxes and borrow heavily. These economic and social strains weakened the dynasty and contributed to its eventual collapse. \\

\addlinespace

\textbf{High Engagement (0.9)} & 
Sure! The Taiping Rebellion (1850–1864) was a massive civil war led by Hong Xiuquan, who claimed to be Jesus’s brother. Fueled by religious vision, social discontent, and political ambition, it drew millions of followers aiming to build a utopian “Heavenly Kingdom.” The rebellion spanned much of southern and central China and inflicted massive casualties. Though defeated, it shook the Qing Dynasty to its core. & 
The rebellion proposed radical social reforms: communal wealth, gender equality, and a new Christian order. Women served in leadership roles, and Confucian hierarchies were challenged. Economically, the war caused immense destruction, crop failure, and trade collapse. Foreign influence increased afterward as China opened to trade and industry to recover. The Taiping Rebellion marked a turning point in modern Chinese history. \\
\bottomrule
\end{tabularx}
\end{table}
\clearpage

\section{Quiz questions and essay prompts}
\label{app:quiz-essay-questions}

\subsection*{T. rex}

\subsubsection*{Fill-in-the-blank Questions}

\begin{enumerate}
    \item \textbf{Q1:} T. rex fossils have primarily been found in \underline{\hspace{1cm}}.  
    (2 points): \textbf{North America}
    
    \item \textbf{Q2:} T. Rex likely obtained its food from hunting as well as \underline{\hspace{1cm}}.  
    (2 points): \textbf{Scavenging}
    
    \item \textbf{Q3:} One of the most complete T. rex fossils, known as \underline{\hspace{1cm}}, was discovered in South Dakota in 1990.  
    (2 points): \textbf{Sue}
\end{enumerate}

\subsubsection*{Multiple Choice Questions}
\textbf{Q4:} Which era did the Tyrannosaurus Rex live in?  
\begin{itemize}
    \item Jurassic
    \item Cretaceous
    \item Triassic
    \item Permian
    \item Devonian
    \item This did not come up in my conversation
    \item I don’t know
\end{itemize}
(2 points): \textbf{Cretaceous}

\textbf{Q5:} How many fingers did the Tyrannosaurus Rex have on each hand?  
\begin{itemize}
    \item Two
    \item Three
    \item Four
    \item Five
    \item Six
    \item This did not come up in my conversation
    \item I don’t know
\end{itemize}
(2 points): \textbf{Two}

\textbf{Q6:} Choose all popular myths about T. Rex that are likely wrong.  
\begin{itemize}
    \item It had poor vision.  
    \item It was a slow, clumsy mover.  
    \item Its arms were likely useless.  
    \item It had one of the most powerful bites ever known.  
    \item It was the dominant dinosaur in its environment.  
    \item It may have had feathers.  
\end{itemize}
(6 points): Correct answers are the first three options: \textbf{It had poor vision}; \textbf{It was a slow, clumsy mover}; \textbf{Its arms were likely useless}.  
For partial credit, 1 point was awarded for each correct choice, and 1 point for not selecting each incorrect choice.

\subsubsection*{Ranking Question}

\textbf{Q9:} Sort the T. Rex into this list of dinosaurs by length.
\begin{enumerate}
    \item Argentinosaurus - 98 ft (30 m)
    \item Brachiosaurus - 82 ft (25 m)
    \item Spinosaurus - 59 ft (18 m)
    \item Iguanodon - 33 ft (10 m)
    \item Pachycephalosaurus - 16 ft (5 m)
    \item Velociraptor - 7 ft (2.5 m)
    \item Tyrannosaurus Rex - ?
\end{enumerate}
(2 points): Tyrannosaurus Rex (40 ft) should be placed \textbf{between Iguanodon and Spinosaurus}.  
(1 point): Placement is one off, in position 3 or 5.

\subsubsection*{Essay Prompt}

Answer one of the following in the form of a mini-essay (introduction, main section, conclusion):
\begin{enumerate}
    \item Discuss T. Rex's physical and behavioral characteristics which enabled it to dominate its environment.
    \item Analyze paleobiological discoveries that have changed our understanding of T. Rex.
    \item Evaluate the theories regarding the function of T. Rex's small arms. What are some of the proposed explanations, and which do you find most convincing?
    \item Create your own essay question related to the T. Rex based on your conversation.
\end{enumerate}

\subsection*{Taiping Rebellion}

\subsubsection*{Fill-in-the-gap Questions}

\begin{enumerate}
    \item \textbf{Q1:} The goal of the Taiping Rebellion was to establish the \underline{\hspace{1cm}}.  
    (2 points): \textbf{Taiping Heavenly Kingdom of Great Peace}, or similar.
    
    \item \textbf{Q2:} One of the distinctive aspects of the Taiping ideology was its spiritual blend of \underline{\hspace{1cm}} and \underline{\hspace{1cm}}, which appealed to the disaffected rural populace.  
    (2 points): \textbf{Christianity} (1) and \textbf{Chinese spiritual traditions} (Buddhism, Taoism, Confucianism) or similar (1).
\end{enumerate}

\subsubsection*{Multiple Choice Questions}

\textbf{Q5:} What was the name of the leader of the Taiping Rebellion?  
\begin{itemize}
    \item Zeng Guofan
    \item Hong Xiuquan
    \item Hong Tianguifu
    \item Feng Yunshan
    \item Yang Xiuqing
    \item This did not come up in my conversation
    \item I don't know
\end{itemize}
(1 point): \textbf{Hong Xiuquan}

\textbf{Q6:} The Qing government was administered by leaders of which ethnic minority?  
\begin{itemize}
    \item Manchu
    \item Hakka
    \item Han
    \item Zhuang
    \item Hui
    \item This did not come up in my conversation
    \item I don't know
\end{itemize}
(1 point): \textbf{Manchu}

\subsubsection*{Ranking Question}

\textbf{Q10:} Here is a list of death tolls in wars by lowest estimate. Where would the Taiping Rebellion sit?  
\begin{enumerate}
    \item World War II - 80 million
    \item World War I - 17 million
    \item Spanish conquest of Mexico - 10.5 million
    \item Russian Civil War - 7 million
    \item Napoleonic Wars - 3.5 million
    \item Vietnam War - 1.3 million
    \item Taiping Rebellion
\end{enumerate}
(2 points): \textbf{Between World War II and World War I}, with an estimated toll of 20-30 million.

\subsubsection*{Essay Prompt}

Answer one of the following in the form of an essay (introduction, main section, conclusion):
\begin{enumerate}
    \item Discuss the socio-economic factors that contributed to the outbreak of the Taiping Rebellion.
    \item Analyze the role of religion in the Taiping Rebellion.
    \item Discuss the legacy of the Taiping Rebellion on subsequent Chinese history.
    \item Create your own essay question related to the Taiping Rebellion based on your conversation.
\end{enumerate}

\end{document}